\documentclass{aa}
\usepackage[varg]{txfonts}
\usepackage{lscape}
\usepackage{graphicx}

\usepackage[table,usenames,dvipsnames]{xcolor}
\definecolor{linkcolor}{rgb}{0.6,0,0}
\definecolor{citecolor}{rgb}{0,0,0.75}
\definecolor{urlcolor}{rgb}{0.12,0.46,0.7}
\usepackage[breaklinks, colorlinks, urlcolor=urlcolor, linkcolor=linkcolor, citecolor=citecolor, pdfencoding=auto]{hyperref}

\def\arcm{\ifmmode {^{\scriptscriptstyle\prime}}
          \else $^{\scriptscriptstyle\prime}$\fi}
\def\pdeg{\ifmmode $\setbox0=\hbox{$^{\circ}$}\rlap{\hskip.11\wd0 .}$^{\circ}
          \else \setbox0=\hbox{$^{\circ}$}\rlap{\hskip.11\wd0 .}$^{\circ}$\fi}

\graphicspath{ {./figures/} }

\def\setsymbol#1#2{\expandafter\def\csname #1\endcsname{#2}}
\def\getsymbol#1{\csname #1\endcsname}

\def\Planck{\textit{Planck}}

\newbox\tablebox    \newdimen\tablewidth
\def\leaderfil{\leaders\hbox to 5pt{\hss.\hss}\hfil}
\def\tablenote#1 #2\par{\begingroup \parindent=0.8em
    \abovedisplayshortskip=0pt\belowdisplayshortskip=0pt
    \noindent
    $$\hss\vbox{\hsize\tablewidth \hangindent=\parindent \hangafter=1 \noindent
    \hbox to \parindent{$^#1$\hss}\strut#2\strut\par}\hss$$
    \endgroup}

\def\L2{\ifmmode L_2\else $L_2$\fi}

\def\DeltaT{\ifmmode \Delta T\else $\Delta T$\fi}
\def\deltat{\ifmmode \Delta t\else $\Delta t$\fi}
\def\fknee{\ifmmode f_{\rm knee}\else $f_{\rm knee}$\fi}
\def\Fmax{\ifmmode F_{\rm max}\else $F_{\rm max}$\fi}
\def\solar{\ifmmode{\rm M}_{\mathord\odot}\else${\rm M}_{\mathord\odot}$\fi}
\def\Msolar{\ifmmode{\rm M}_{\mathord\odot}\else${\rm M}_{\mathord\odot}$\fi}
\def\Lsolar{\ifmmode{\rm L}_{\mathord\odot}\else${\rm L}_{\mathord\odot}$\fi}
\def\inv{\ifmmode^{-1}\else$^{-1}$\fi}
\def\mo{\ifmmode^{-1}\else$^{-1}$\fi}
\def\sup#1{\ifmmode ^{\rm #1}\else $^{\rm #1}$\fi}
\def\expo#1{\ifmmode \times 10^{#1}\else $\times 10^{#1}$\fi}
\def\,{\thinspace}
\def\lsim{\mathrel{\raise .4ex\hbox{\rlap{$<$}\lower 1.2ex\hbox{$\sim$}}}}
\def\gsim{\mathrel{\raise .4ex\hbox{\rlap{$>$}\lower 1.2ex\hbox{$\sim$}}}}

\def\simprop{\mathrel{\raise .4ex\hbox{\rlap{$\propto$}\lower 1.2ex\hbox{$\sim$}}}}
\def\deg{\ifmmode^\circ\else$^\circ$\fi}
\def\pdeg{\ifmmode $\setbox0=\hbox{$^{\circ}$}\rlap{\hskip.11\wd0 .}$^{\circ}
          \else \setbox0=\hbox{$^{\circ}$}\rlap{\hskip.11\wd0 .}$^{\circ}$\fi}
\def\arcs{\ifmmode {^{\scriptstyle\prime\prime}}
          \else $^{\scriptstyle\prime\prime}$\fi}
\def\arcm{\ifmmode {^{\scriptstyle\prime}}
          \else $^{\scriptstyle\prime}$\fi}
\newdimen\sa  \newdimen\sb
\def\parcs{\sa=.07em \sb=.03em
     \ifmmode \hbox{\rlap{.}}^{\scriptstyle\prime\kern -\sb\prime}\hbox{\kern -\sa}
     \else \rlap{.}$^{\scriptstyle\prime\kern -\sb\prime}$\kern -\sa\fi}
\def\parcm{\sa=.08em \sb=.03em
     \ifmmode \hbox{\rlap{.}\kern\sa}^{\scriptstyle\prime}\hbox{\kern-\sb}
     \else \rlap{.}\kern\sa$^{\scriptstyle\prime}$\kern-\sb\fi}
\def\ra[#1 #2 #3.#4]{#1\sup{h}#2\sup{m}#3\sup{s}\llap.#4}
\def\dec[#1 #2 #3.#4]{#1\deg#2\arcm#3\arcs\llap.#4}
\def\deco[#1 #2 #3]{#1\deg#2\arcm#3\arcs}
\def\rra[#1 #2]{#1\sup{h}#2\sup{m}}

\def\dots{\relax\ifmmode \ldots\else $\ldots$\fi}
\def\WHzsr{\ifmmode $W\,Hz\mo\,sr\mo$\else W\,Hz\mo\,sr\mo\fi}
\def\mHz{\ifmmode $\,mHz$\else \,mHz\fi}
\def\GHz{\ifmmode $\,GHz$\else \,GHz\fi}
\def\mKs{\ifmmode $\,mK\,s$^{1/2}\else \,mK\,s$^{1/2}$\fi}
\def\muKs{\ifmmode \,\mu$K\,s$^{1/2}\else \,$\mu$K\,s$^{1/2}$\fi}
\def\muKRJs{\ifmmode \,\mu$K$_{\rm RJ}$\,s$^{1/2}\else \,$\mu$K$_{\rm RJ}$\,s$^{1/2}$\fi}
\def\muKHz{\ifmmode \,\mu$K\,Hz$^{-1/2}\else \,$\mu$K\,Hz$^{-1/2}$\fi}
\def\MJysr{\ifmmode \,$MJy\,sr\mo$\else \,MJy\,sr\mo\fi}
\def\MJysrmK{\ifmmode \,$MJy\,sr\mo$\,mK$_{\rm CMB}\mo\else \,MJy\,sr\mo\,mK$_{\rm CMB}\mo$\fi}
\def\microns{\ifmmode \,\mu$m$\else \,$\mu$m\fi}

\def\muK{\ifmmode \,\mu$K$\else \,$\mu$\hbox{K}\fi}
\def\microK{\ifmmode \,\mu$K$\else \,$\mu$\hbox{K}\fi}
\def\muW{\ifmmode \,\mu$W$\else \,$\mu$\hbox{W}\fi}
\def\kms{\ifmmode $\,km\,s$^{-1}\else \,km\,s$^{-1}$\fi}
\def\kmsMpc{\ifmmode $\,\kms\,Mpc\mo$\else \,\kms\,Mpc\mo\fi}
\providecommand{\sorthelp}[1]{}

\begin{document}

\title{Polarization and variability of compact sources measured in \textit{Planck}  time-ordered data}
\author{
G.~Rocha\inst{1, 2}
\and
R.~Keskitalo\inst{3}
\and
B.~Partridge\inst{4}
\and
A.~Marscher\inst{5}
\and
C.~O'Dea\inst{6}
\and
T.~J.~Pearson\inst{2,7}
\and
K.~M.~G\'{o}rski\inst{1, 8}
}
\institute{
Jet Propulsion Laboratory, California Institute of Technology, 4800 Oak Grove Drive, Pasadena, California, 91109, U.S.A.\goodbreak
\and
California Institute of Technology, 1200 E California Blvd, Pasadena, California, 91125, U.S.A.\goodbreak
\and
Computational Cosmology Center, Lawrence Berkeley National Laboratory, Berkeley, California, U.S.A.\goodbreak
\and
Haverford College Astronomy Department, 370 Lancaster Avenue, Haverford, Pennsylvania, U.S.A.\goodbreak
\and
Institute for Astrophysical Research, Boston University, 725 Commonwealth Ave., Boston, MA 02215, U.S.A.\goodbreak
\and
University Of Manitoba, Winnipeg, Manitoba, Canada \goodbreak
\and
Infrared Processing and Analysis Center, California Institute of Technology, Pasadena, CA 91125, U.S.A.\goodbreak
\and
Warsaw University Observatory, Aleje Ujazdowskie 4, 00-478 Warszawa, Poland\goodbreak
}
\date{June 2022}

\abstract{
This paper introduces a new \Planck\ Catalog of Polarized and Variable Compact Sources (PCCS-PV) comprising 153 sources, the majority of which are extragalactic. The data include both the total flux density and linear polarization measured by \Planck\, with frequency coverage from 30 to 353 GHz, and temporal spacing ranging from days to years.
We classify most sources as beamed, extragalactic radio sources; the catalog also includes several radio galaxies, Seyfert galaxies, and Galactic and Magellanic Cloud sources,  including \ion{H}{ii} regions and planetary nebulae.
An advanced  extraction method applied directly to the multifrequency Planck time-ordered data, rather than the mission sky maps, was developed to allow an assessment of the variability of polarized sources.
Our analysis of the time-ordered data from the \Planck\ mission, tod2flux, allowed us to catalog the time-varying emission and polarization properties for these sources at the full range of polarized frequencies employed by \Planck, 30 to 353 GHz. PCCS-PV provides the time- and frequency-dependent, polarized flux densities for all 153 sources.
To illustrate some potential applications of the PCCS-PV, we conducted preliminary comparisons of  our measurements of selected sources with published data from other astronomical instruments. In summary, we find general agreement between the \Planck\ and the Institut de Radioastronomie Millim{\'e}trique (IRAM) polarization measurements as well as with the Metsähovi 37\,GHz values at closely similar epochs. These combined measurements also show the value of PCCS-PV results and the PCCS2 catalog for filling in missing spectral (or temporal) coverage and helping to define the spectral energy distributions of extragalactic sources.  In turn, these results provide useful clues as to the physical properties of the sources.}


\titlerunning{Polarized variable compact sources from \textit{Planck} TOD}
\authorrunning{Rocha et al.}

\maketitle

\section{Introduction}
\label{sec:introduction}
There are many categories of prominent radio, millimeter-wave, and submillimeter sources, including Galactic sources such as \ion{H}{ii} regions and supernovae remnants, and extragalactic systems such as quasars and dusty star-forming galaxies. The brightest of these sources at high frequencies, however, are typically either Galactic or active galactic nuclei (AGN). Both classes are considered here, though the bulk of our objects are extragalactic. AGN are the most powerful long-lasting sources in the Universe. Unlike much shorter timescale events such as black hole mergers or gamma-ray bursts, the active phase lasts long enough so that its properties and evolution can be studied in detail. Radio and submillimeter observations are central to understanding these systems.  The radio emission of AGN is almost certainly related to a black hole at the center of the active galaxy and to the jets it engenders.  These massive black holes in turn appear to be intimately related to the evolution of their host galaxies \citep{Magorrian1998}.   AGN feedback in the form of radiation, winds, and relativistic jets is now invoked to explain the galaxy luminosity function and the $M$--$\sigma$ relation \citep[e.g.,][]{Croton2006,Fabian2012, King2015}.  The total power in the jets is a critical parameter in estimating the magnitude of this feedback.  The physics required to estimate the jet power, however, is poorly understood.  Important clues are provided by studies of both polarization and variability of emission from the jets. Such observations can illuminate the launching and initial propagation of relativistic jets, as well as the presence of jet ``sheaths'' and surrounding winds that could contribute to feedback \citep{Blandford2019}.  The optical depth of the material surrounding the central black hole increases with wavelength, however.  Hence millimeter and submillimeter observations are particularly useful: they allow us to study the properties of both the black hole and its surrounding accretion disk, as has been done for a few bright, relatively nearby, radio galaxies such as M87 (3C\,274) \citep{Junor1999,Hada2011}  and 3C\,84 (NGC\,1275)
\citep{Giovannini2018}. In these objects, as well as more distant radio galaxies and blazars, the millimeter--submillimeter continuum spectrum, time variability, and linear polarization probe the acceleration and collimation zone of the jet.
  
The observed properties of AGN strongly depend on the orientation between the line of sight and the axis of the jets.  Emission from the jet itself is dominant when the line of sight and jet axis align closely. This class of AGN, the blazars, includes BL Lac objects and flat-spectrum radio quasars (FSRQs). Doppler boosting causes the emission of the aligned jet to be strong and the counter-jet to be weak.  A substantial fraction of the AGN detected by \Planck\ and other cosmic microwave background (CMB) experiments are blazars \citep{planck2011-6.1}. 
That is not surprising, given the high apparent luminosity of blazars and their continuum spectra, which peak at millimeter to submillimeter wavelengths.  The close alignment of jet and line of sight also allows for variability introduced by small changes in the angle between the line of sight and the jet \citep{Begelman1984,Chen2013}. 

To maximize the science that can be extracted from radio and submillimeter observations of AGN, we ideally need high angular resolution, wide frequency coverage (extending to the submillimeter range) and repeated observations over time-scales of days to years including both total power and polarization measurements. Ground based interferometers, including the Atacama Large Millimeter Array (ALMA), meet the first of these needs.  We show here that data derived as a byproduct of  CMB experiments, properly analysed, can meet the last two.  The sensitivity of current CMB experiments, including the \Planck\ mission, is adequate to detect thousands of the brightest AGN. Most current CMB experiments now employ polarization-sensitive detectors and combine observations at several frequencies. \Planck, for instance made polarized observations of the entire sky at seven frequencies from 30 to 353\,GHz, some of them inaccessible from the ground. Finally, CMB searches generally cover large areas of the sky (the entire sky, in the case of \Planck) repeatedly, providing observations of sources with cadences ranging from minutes to years.
 
To date, however, the potential contributions of CMB experiments and data sets have not been fully realized because the repeated observations of a given sky region are merely added together to produce static sky maps \citep{Partridge2017}. Combining repeated observations such as this makes perfect sense for the CMB and for Galactic foregrounds, both of which are static, but all information in the time domain is lost.  We remedy that situation here by working with the time-ordered data (ToD) available from the multiyear \Planck\ mission. The methods we employ can easily be extended to data coming from other CMB surveys.  Here we provide a catalog of extragalactic sources measured in both total power and polarization by \Planck, with frequency coverage from 30 to 353\,GHz, and time-scales ranging from days to years.

We discuss the \Planck\ mission, source selection, calibration and related topics in Sect.~\ref{sec:planck}.  In Sect.~\ref{sec:method}, we describe our technical approach to extracting flux density and polarization measurements from the \Planck\ ToD, source by source and epoch by epoch.  Sect.~\ref{sec:catalog} describes the catalog we construct of these observations.  Sect.~\ref{sec:results} then treats the results and some consequences for models of AGN.  We conclude in  Sect.~\ref{sec:summary}.

\section{\Planck\ data}
\label{sec:planck}

All the data employed here were taken from archived time-ordered data (ToD) from the \Planck\ mission as found in the Planck Legacy Archive.\footnote{\url{https://www.cosmos.esa.int/web/planck/pla}}  We first describe aspects of the mission relevant to our study, then describe how we selected bright sources.  While most bright sources are AGN, we include some Galactic and thermal sources in our study.  Galactic sources, such as the planetary nebula NGC\,6572, are not expected to vary on short time scales, and hence provide a useful test of our pipeline.  

\subsection{The \Planck\ mission}

\Planck\ was a European Space Agency (ESA) mission designed to map fluctuations in the CMB to the limit set by unavoidable cosmic variance down to angular
scales of 5' or so \citep{planck2016-l01}.
Two instruments shared a common focal plane: the low-frequency instrument (LFI), with HEMT radiometers operating at 30, 44 and 70\,GHz, and the high-frequency instrument employing lower-noise bolometric detectors in six frequency bands from 100 to 857\,GHz.  The fractional bandwidth at each frequency ranged from 20\% to 30\%.  Polarization was measured at all but the two highest-frequency bands (545 and 857\,GHz).  Table~\ref{tab:planck} lists some of the relevant properties of the \Planck\ mission.  While mapping the CMB, \Planck\ detected thousands of compact sources, both Galactic and extragalactic.  \Planck\ thus provides a valuable archive of observations of these sources, which we explore in this paper.  These sources are cataloged, frequency by frequency, in several \Planck\ catalogs; we employ only the latest, PCCS2 \citep{planck2014-a35}.
Approximate completeness limits in this catalog are provided in Table~\ref{tab:planck}.  These limits give  a rough indication of the sensitivity of \Planck\ to compact (unresolved) sources in each frequency band. A multifrequency catalog of \Planck\ nonthermal sources is provided in \citet{planck2018-LIV}. 

The scan strategy employed by \Planck\ ensured that the full sky was covered twice in one year \citep{planck2013-p01}.
Typically, a given source was observed at a given \Planck\ frequency for a few days to one week in  each six month survey (sources near the ecliptic poles, where \Planck\ scans overlapped, were viewed much more frequently).  HFI completed approximately five full sky surveys from 2009 August to 2012 January before its coolant ran out. LFI completed a fraction more than eight six-month surveys stretching from 2009 August to 2013 October.  Thus we have five independent samples of flux densities at nine frequencies for each source, and an additional three or four measurements at 30, 44 and 70\,GHz.  The flux densities recorded in PCCS2 are averages over the entire mission, however, and thus smooth out any variability.  In this paper, we work with the independent samples by analysing the time-ordered data (ToD).

\Planck\ calibration at the frequencies employed here is determined from the satellite's annual motion around the solar system barycenter 
(\citealt{planck2016-l02} for LFI; \citealt{planck2016-l03} for HFI).
The  calibration is thus absolute and precise to subpercent accuracy.  This method of calibration has two consequences.  First, the annual motion of the satellite produces a dipole signal in the CMB, much larger in angular scale than the beam width.  In contrast the compact sources we investigate are almost never resolved, even at \Planck's highest frequency.  Accurate calculations of flux density therefore require exact knowledge of  the beam solid angle.  The beam solid angles are known at each \Planck\ band to percent-level accuracy \citep{planck2016-l01}. 
Second, the substantial bandwidth of the \Planck\ detectors means that they respond differently to sources with spectra that differ from the blackbody spectrum of the CMB.  Thus, for precision work,  each measured flux density needs a small, spectrum-dependent, ``color correction.''  We  return to this issue in Sect.~\ref{sec:catalog} below.

\begin{table*}[ht!]
  \newcommand\ph{\phantom{0}}
  \begin{center}
    \caption{Relevant properties of the \Planck\ instruments. \textbf{}}
    \label{tab:planck}
    \begin{tabular}{l@{\,\,\dots\,\dots\,\dots\qquad} c c c c}
    \hline\hline
      \multicolumn{1}{c}{Band frequency}\hfill&Center frequency\tablefootmark{a}&FWHM&Polarized?&90\% completeness\\
      \multicolumn{1}{c}{(GHz)}&(GHz)&(arcmin)&&(mJy)\\
      \hline
      \ph{}30&  \ph{}28.3&   32.29& yes& 430\\
      \ph{}44&  \ph44.0&   27.94& yes& 700\\
      \ph70&  \ph{}70.2&   13.08& yes& 500\\
      100&  100.3&   \ph9.66& yes& 270\\
      143&  141.6&   \ph7.22& yes& 180\\
      217&  219.6&   \ph4.90& yes& 150\\
      353&  359.0&   \ph4.92& yes& 300\\
      545&  555.4&   \ph4.67& no& 550\\
      857&  850.0&   \ph4.22& no& 790\\
      \hline
    \end{tabular}
  \end{center}
  \tablefoot{
  \tablefoottext{a}{The center frequencies are calculated for sources with synchrotron spectra.}}
\end{table*}

\begin{figure}
    \centering
    \includegraphics[width=\columnwidth]{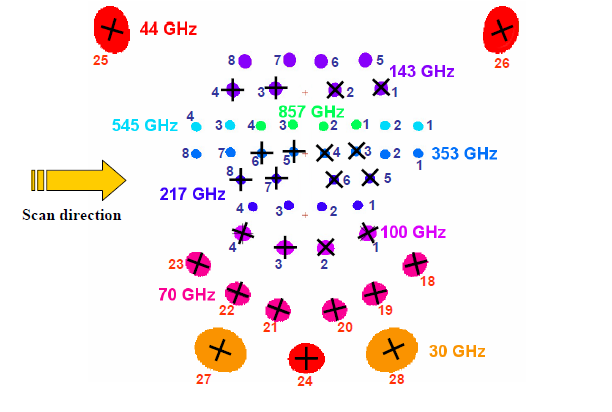}
    \caption{Footprint of the \Planck\ focal plane as seen by an observer at infinity.  The size of each colored spot is a rough indication of the relative resolution.  The small crosses indicate polarization sensitivity.}
    \label{fig:my_label}
\end{figure}

\subsection{Source selection}

\begin{figure}
    \centering
    \includegraphics[width=\columnwidth,trim=0 45 0 45,clip]{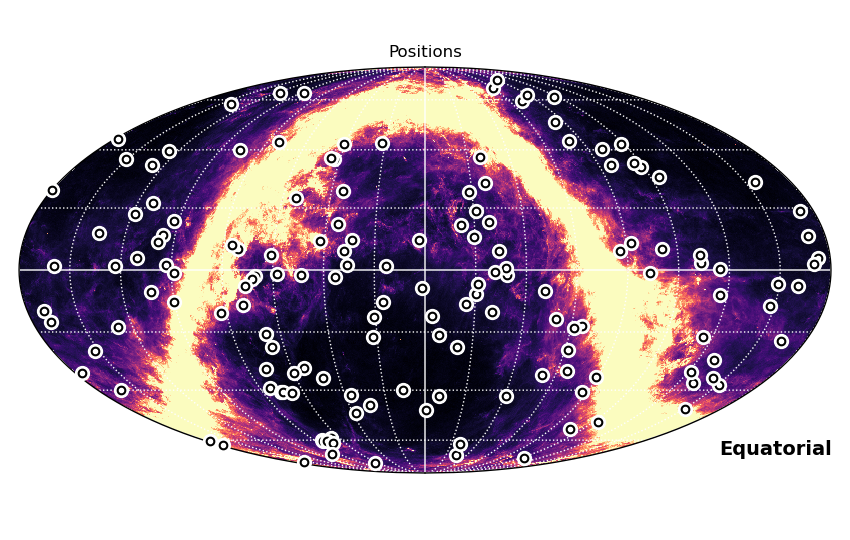}
    \caption{Positions of all 153 sources plotted over the \Planck\ 857\,GHz intensity map.}
    \label{fig:positions}
\end{figure}

Given \Planck's relatively low sensitivity to compact sources and the low polarization fraction expected (a few percent; see, e.g., \citealt{Datta2019}), we elected to work only with the brightest sources in the \Planck\  PCCS2 catalog \citep{planck2014-a35}.  
Specifically, we selected all PCCS2 sources with 100\,GHz flux density greater than 1\,Jy (roughly 15--20 times the $1\sigma$ uncertainty).  Recall that the cataloged \Planck\ flux densities are averages over the entire duration of the mission.
Since most of these bright sources are variable, some may have fallen below the 1\,Jy threshold for part of that period.  The PCCS2 catalog yielded 148 sources above the 1\,Jy threshold at 100\,GHz; 111 were at Galactic latitude $>20\deg$ and therefore likely to be extragalactic.  This list included many well-known AGN such as 3C\,454.3 and BL\,Lac.  To this initial list, we added five sources well-studied in other high-frequency monitoring programs.  In the end, we examined 153 sources, both Galactic and extragalactic.  The majority were extragalactic, and most of these were blazars.  
The full list of sources with classification is shown in Table~\ref{tab:targets}.  Positions of all of the sources are indicated in Fig.~\ref{fig:positions}.

\longtab{
  \begin{longtable}{lllrr}
    \caption{\label{tab:targets} All fitted sources.} \\
    \hline\hline
    PCCS2 name & Alias & Type & RA & Dec \\
    \hline
    \endfirsthead
    \caption{(continued.)} \\
    \hline\hline
    PCCS2 name & Alias & Type & RA & Dec \\
    \hline
    \endhead
    \hline
    \endfoot
    PCCS2 143 G000.35$-$19.48 & PGCC G000.37$-$19.50 & Cold clump & $19^\mathrm h\,10^\mathrm m\,16^\mathrm s$ & $-37\deg 08\arcm 52\arcs$ \\
PCCS2 143 G001.36+45.99 & ICRF J151640.2+001501 & Seyfert 1 Galaxy & $15^\mathrm h\,16^\mathrm m\,36^\mathrm s$ & $+00\deg 14\arcm 38\arcs$ \\
PCCS2 143 G001.58$-$28.95 & QSO B1954$-$388 & Quasar & $19^\mathrm h\,57^\mathrm m\,58^\mathrm s$ & $-38\deg 44\arcm 56\arcs$ \\
PCCS2 143 G009.33$-$19.61 & QSO B1921$-$293 & BL Lac - type object & $19^\mathrm h\,24^\mathrm m\,52^\mathrm s$ & $-29\deg 15\arcm 26\arcs$ \\
PCCS2 143 G010.83+40.91 & QSO B1546+0246 & Quasar & $15^\mathrm h\,49^\mathrm m\,27^\mathrm s$ & $+02\deg 36\arcm 56\arcs$ \\
PCCS2 143 G014.20+42.21 & 4C 05.64 & Quasar & $15^\mathrm h\,50^\mathrm m\,36^\mathrm s$ & $+05\deg 26\arcm 20\arcs$ \\
PCCS2 143 G016.86$-$13.20 & QSO B1908$-$202 & Quasar & $19^\mathrm h\,11^\mathrm m\,05^\mathrm s$ & $-20\deg 06\arcm 58\arcs$ \\
PCCS2 143 G017.19$-$16.26 & PMN J1923$-$2104 & Quasar & $19^\mathrm h\,23^\mathrm m\,35^\mathrm s$ & $-21\deg 03\arcm 53\arcs$ \\
PCCS2 143 G021.20+19.62 & 3C 353 & Radio Galaxy & $17^\mathrm h\,20^\mathrm m\,32^\mathrm s$ & $-00\deg 59\arcm 08\arcs$ \\
PCCS2 143 G024.01$-$23.11 & QSO B1958-179 & Quasar & $20^\mathrm h\,00^\mathrm m\,57^\mathrm s$ & $-17\deg 48\arcm 38\arcs$ \\
PCCS2 143 G024.37$-$64.93 & ICRF J225805.9$-$275821 & Seyfert 1 Galaxy & $22^\mathrm h\,58^\mathrm m\,08^\mathrm s$ & $-27\deg 58\arcm 45\arcs$ \\
PCCS2 143 G026.68+28.66 & QSO J1658+0741 & Quasar & $16^\mathrm h\,58^\mathrm m\,05^\mathrm s$ & $+07\deg 41\arcm 22\arcs$ \\
PCCS2 143 G034.61+11.84 & NGC 6572 & Planetary Nebula & $18^\mathrm h\,12^\mathrm m\,07^\mathrm s$ & $+06\deg 50\arcm 50\arcs$ \\
PCCS2 143 G034.90+17.65 & QSO B1749+096 & BL Lac - type object & $17^\mathrm h\,51^\mathrm m\,30^\mathrm s$ & $+09\deg 38\arcm 34\arcs$ \\
PCCS2 143 G036.89$-$24.37 & QSO B2022$-$077 & BL Lac - type object & $20^\mathrm h\,25^\mathrm m\,38^\mathrm s$ & $-07\deg 35\arcm 43\arcs$ \\
PCCS2 143 G039.59$-$72.18 & ICRF J233355.2$-$234340 & BL Lac - type object & $23^\mathrm h\,33^\mathrm m\,52^\mathrm s$ & $-23\deg 44\arcm 03\arcs$ \\
PCCS2 143 G040.66$-$48.04 & QSO B2155$-$152 & Quasar & $21^\mathrm h\,58^\mathrm m\,12^\mathrm s$ & $-15\deg 00\arcm 44\arcs$ \\
PCCS2 143 G052.40$-$36.49 & 4C $-$02.81 & BL Lac - type object & $21^\mathrm h\,34^\mathrm m\,10^\mathrm s$ & $-01\deg 52\arcm 39\arcs$ \\
PCCS2 143 G053.89$-$57.06 & QSO J2246$-$1206 & Quasar & $22^\mathrm h\,46^\mathrm m\,18^\mathrm s$ & $-12\deg 06\arcm 01\arcs$ \\
PCCS2 143 G055.14+46.38 & QSO B1611+343 & Quasar & $16^\mathrm h\,13^\mathrm m\,39^\mathrm s$ & $+34\deg 12\arcm 28\arcs$ \\
PCCS2 143 G055.22$-$51.71 & QSO B2227$-$0848 & Quasar & $22^\mathrm h\,29^\mathrm m\,42^\mathrm s$ & $-08\deg 33\arcm 05\arcs$ \\
PCCS2 143 G055.47$-$35.57 & QSO B2134+0028 & Quasar & $21^\mathrm h\,36^\mathrm m\,38^\mathrm s$ & $+00\deg 42\arcm 09\arcs$ \\
PCCS2 143 G057.68$-$40.34 & ICRF J215614.7$-$003704 & BL Lac - type object & $21^\mathrm h\,56^\mathrm m\,13^\mathrm s$ & $-00\deg 38\arcm 11\arcs$ \\
PCCS2 143 G058.95$-$48.84 & 3C 446 & Quasar & $22^\mathrm h\,25^\mathrm m\,47^\mathrm s$ & $-04\deg 57\arcm 14\arcs$ \\
PCCS2 143 G061.08+42.34 & QSO B1633+382 & Quasar & $16^\mathrm h\,35^\mathrm m\,13^\mathrm s$ & $+38\deg 08\arcm 14\arcs$ \\
PCCS2 143 G063.45+40.94 & 3C 345 & Quasar & $16^\mathrm h\,43^\mathrm m\,01^\mathrm s$ & $+39\deg 48\arcm 40\arcs$ \\
PCCS2 143 G063.66$-$34.06 & 4C 06.69 & Quasar & $21^\mathrm h\,48^\mathrm m\,06^\mathrm s$ & $+06\deg 58\arcm 05\arcs$ \\
PCCS2 143 G064.03+31.01 & QSO J1734+3857 & BL Lac - type object & $17^\mathrm h\,34^\mathrm m\,18^\mathrm s$ & $+38\deg 58\arcm 19\arcs$ \\
PCCS2 143 G065.54$-$71.88 & QSO B2345$-$167 & Quasar & $23^\mathrm h\,48^\mathrm m\,01^\mathrm s$ & $-16\deg 30\arcm 47\arcs$ \\
PCCS2 143 G071.41+33.28 & ICRF J172727.6+453039 & Seyfert 1 Galaxy & $17^\mathrm h\,27^\mathrm m\,27^\mathrm s$ & $+45\deg 29\arcm 59\arcs$ \\
PCCS2 143 G073.45+41.87 & 4C 47.44 & Quasar & $16^\mathrm h\,37^\mathrm m\,45^\mathrm s$ & $+47\deg 18\arcm 11\arcs$ \\
PCCS2 143 G075.66$-$29.63 & QSO B2201+1711 & BL Lac - type object & $22^\mathrm h\,03^\mathrm m\,25^\mathrm s$ & $+17\deg 25\arcm 45\arcs$ \\
PCCS2 143 G077.22+23.48 & ICRF J182931.7+484446 & Seyfert 1 Galaxy & $18^\mathrm h\,29^\mathrm m\,36^\mathrm s$ & $+48\deg 44\arcm 19\arcs$ \\
PCCS2 143 G077.42$-$38.59 & CTA102 & Quasar & $22^\mathrm h\,32^\mathrm m\,35^\mathrm s$ & $+11\deg 43\arcm 10\arcs$ \\
PCCS2 143 G083.16$-$30.08 & QSO J2225+2118 & Quasar & $22^\mathrm h\,25^\mathrm m\,38^\mathrm s$ & $+21\deg 18\arcm 35\arcs$ \\
PCCS2 143 G085.67+83.35 & QSO B1308+326 & Quasar & $13^\mathrm h\,10^\mathrm m\,27^\mathrm s$ & $+32\deg 20\arcm 06\arcs$ \\
PCCS2 143 G085.73+26.08 & 4C 56.27 & BL Lac - type object & $18^\mathrm h\,24^\mathrm m\,03^\mathrm s$ & $+56\deg 50\arcm 51\arcs$ \\
PCCS2 143 G085.95$-$18.78 & 4C 31.63 & BL Lac - type object & $22^\mathrm h\,03^\mathrm m\,16^\mathrm s$ & $+31\deg 45\arcm 36\arcs$ \\
PCCS2 143 G086.11$-$38.19 & 3C 454.3 & Quasar & $22^\mathrm h\,53^\mathrm m\,58^\mathrm s$ & $+16\deg 08\arcm 34\arcs$ \\
PCCS2 143 G090.10$-$25.64 & QSO B2232+282 & BL Lac - type object & $22^\mathrm h\,36^\mathrm m\,20^\mathrm s$ & $+28\deg 28\arcm 37\arcs$ \\
PCCS2 143 G092.58$-$10.44 & BL Lac & BL Lac - type object & $22^\mathrm h\,02^\mathrm m\,43^\mathrm s$ & $+42\deg 16\arcm 43\arcs$ \\
PCCS2 143 G093.48$-$66.64 & QSO B0003$-$066 & BL Lac - type object & $00^\mathrm h\,06^\mathrm m\,12^\mathrm s$ & $-06\deg 23\arcm 32\arcs$ \\
PCCS2 143 G097.47+25.03 & ICRF J184916.0+670541 & Seyfert 1 Galaxy & $18^\mathrm h\,49^\mathrm m\,18^\mathrm s$ & $+67\deg 04\arcm 57\arcs$ \\
PCCS2 143 G100.13+29.17 & 3C 371 & Radio Galaxy & $18^\mathrm h\,06^\mathrm m\,46^\mathrm s$ & $+69\deg 49\arcm 33\arcs$ \\
PCCS2 143 G100.68+36.61 & 4C 69.21 & Quasar & $16^\mathrm h\,42^\mathrm m\,14^\mathrm s$ & $+68\deg 56\arcm 00\arcs$ \\
PCCS2 143 G105.62+23.54 & ICRF J192748.4+735801 & Seyfert 1 Galaxy & $19^\mathrm h\,27^\mathrm m\,40^\mathrm s$ & $+73\deg 57\arcm 58\arcs$ \\
PCCS2 143 G106.95$-$50.62 & Mrk 1501 & Seyfert 1 Galaxy & $00^\mathrm h\,10^\mathrm m\,27^\mathrm s$ & $+10\deg 58\arcm 26\arcs$ \\
PCCS2 143 G110.03+29.07 & QSO B1803+784 & BL Lac - type object & $18^\mathrm h\,00^\mathrm m\,34^\mathrm s$ & $+78\deg 27\arcm 42\arcs$ \\
PCCS2 143 G130.79$-$14.31 & QSO B0133+476 & Quasar & $01^\mathrm h\,37^\mathrm m\,02^\mathrm s$ & $+47\deg 51\arcm 33\arcs$ \\
PCCS2 143 G131.81$-$60.99 & 4C 01.02 & Quasar & $01^\mathrm h\,08^\mathrm m\,37^\mathrm s$ & $+01\deg 34\arcm 50\arcs$ \\
PCCS2 143 G143.53+34.42 & QSO B0836+710 & Quasar & $08^\mathrm h\,41^\mathrm m\,25^\mathrm s$ & $+70\deg 54\arcm 09\arcs$ \\
PCCS2 143 G143.97+28.02 & QSO B0716+714 & BL Lac - type object & $07^\mathrm h\,21^\mathrm m\,59^\mathrm s$ & $+71\deg 21\arcm 02\arcs$ \\
PCCS2 143 G144.62$-$73.42 & QSO B0113$-$1152 & Quasar & $01^\mathrm h\,16^\mathrm m\,09^\mathrm s$ & $-11\deg 36\arcm 01\arcs$ \\
PCCS2 143 G144.97$-$09.85 & 4C 47.08 & BL Lac - type object & $03^\mathrm h\,03^\mathrm m\,32^\mathrm s$ & $+47\deg 17\arcm 07\arcs$ \\
PCCS2 143 G145.57+64.99 & 4C 49.22 & Quasar & $11^\mathrm h\,53^\mathrm m\,22^\mathrm s$ & $+49\deg 30\arcm 26\arcs$ \\
PCCS2 143 G145.73+43.13 & QSO B0954+658 & BL Lac - type object & $09^\mathrm h\,58^\mathrm m\,53^\mathrm s$ & $+65\deg 34\arcm 06\arcs$ \\
PCCS2 143 G149.46$-$28.52 & QSO B0234+285 & Quasar & $02^\mathrm h\,37^\mathrm m\,52^\mathrm s$ & $+28\deg 48\arcm 02\arcs$ \\
PCCS2 143 G149.56$-$14.10 & ICRF J031301.9+412001 & Seyfert 1 Galaxy & $03^\mathrm h\,12^\mathrm m\,58^\mathrm s$ & $+41\deg 19\arcm 45\arcs$ \\
PCCS2 143 G150.56$-$13.25 & 3C 84 & Radio Galaxy & $03^\mathrm h\,19^\mathrm m\,46^\mathrm s$ & $+41\deg 31\arcm 28\arcs$ \\
PCCS2 143 G152.56$-$47.35 & ICRF J021113.1+105134 & BL Lac - type object & $02^\mathrm h\,11^\mathrm m\,11^\mathrm s$ & $+10\deg 52\arcm 29\arcs$ \\
PCCS2 143 G156.77$-$39.10 & QSO B0235+1624 & BL Lac - type object & $02^\mathrm h\,38^\mathrm m\,40^\mathrm s$ & $+16\deg 37\arcm 05\arcs$ \\
PCCS2 143 G159.70$-$49.11 & 4C 06.11 & Quasar & $02^\mathrm h\,24^\mathrm m\,27^\mathrm s$ & $+06\deg 59\arcm 55\arcs$ \\
PCCS2 143 G162.12$-$54.41 & QSO J0217+0144 & Quasar & $02^\mathrm h\,17^\mathrm m\,47^\mathrm s$ & $+01\deg 45\arcm 12\arcs$ \\
PCCS2 143 G162.15+08.22 & 6C 052928+482031 & Quasar & $05^\mathrm h\,33^\mathrm m\,13^\mathrm s$ & $+48\deg 23\arcm 01\arcs$ \\
PCCS2 143 G168.07$-$19.15 & LDN 1489 & Dark Cloud (nebula) & $04^\mathrm h\,04^\mathrm m\,46^\mathrm s$ & $+26\deg 19\arcm 10\arcs$ \\
PCCS2 143 G168.10$-$76.03 & QSO B0130$-$17 & Quasar & $01^\mathrm h\,32^\mathrm m\,39^\mathrm s$ & $-16\deg 55\arcm 32\arcs$ \\
PCCS2 143 G169.15$-$39.71 & QSO B0306+101 & Quasar & $03^\mathrm h\,09^\mathrm m\,01^\mathrm s$ & $+10\deg 30\arcm 15\arcs$ \\
PCCS2 143 G171.08+17.94 & QSO J0646+4451 & Quasar & $06^\mathrm h\,46^\mathrm m\,30^\mathrm s$ & $+44\deg 51\arcm 37\arcs$ \\
PCCS2 143 G174.12$-$54.30 & QSO B0237$-$027 & Quasar & $02^\mathrm h\,39^\mathrm m\,44^\mathrm s$ & $-02\deg 34\arcm 05\arcs$ \\
PCCS2 143 G175.67+44.82 & QSO B0917+449 & Quasar & $09^\mathrm h\,21^\mathrm m\,02^\mathrm s$ & $+44\deg 42\arcm 47\arcs$ \\
PCCS2 143 G177.39+58.35 & QSO B1030+415 & Quasar & $10^\mathrm h\,33^\mathrm m\,00^\mathrm s$ & $+41\deg 15\arcm 55\arcs$ \\
PCCS2 143 G183.72+46.16 & ICRF J092703.0+390220 & Seyfert 1 Galaxy & $09^\mathrm h\,27^\mathrm m\,02^\mathrm s$ & $+39\deg 01\arcm 32\arcs$ \\
PCCS2 143 G187.98$-$42.44 & CTA 26 & Quasar & $03^\mathrm h\,39^\mathrm m\,30^\mathrm s$ & $-01\deg 45\arcm 36\arcs$ \\
PCCS2 143 G190.36$-$27.39 & 3C 120 & Radio Galaxy & $04^\mathrm h\,33^\mathrm m\,10^\mathrm s$ & $+05\deg 21\arcm 37\arcs$ \\
PCCS2 143 G195.28$-$33.13 & QSO B0420$-$0127 & Quasar & $04^\mathrm h\,23^\mathrm m\,16^\mathrm s$ & $-01\deg 19\arcm 57\arcs$ \\
PCCS2 143 G196.83$-$13.74 & QSO B0529+075 & Quasar & $05^\mathrm h\,32^\mathrm m\,37^\mathrm s$ & $+07\deg 32\arcm 52\arcs$ \\
PCCS2 143 G196.94$-$10.36 & LDN 1594 & Cloud & $05^\mathrm h\,44^\mathrm m\,39^\mathrm s$ & $+09\deg 09\arcm 25\arcs$ \\
PCCS2 143 G199.41+78.37 & QSO B1156+295 & Quasar & $11^\mathrm h\,59^\mathrm m\,32^\mathrm s$ & $+29\deg 14\arcm 45\arcs$ \\
PCCS2 143 G200.02+31.87 & QSO B0827+243 & Quasar & $08^\mathrm h\,30^\mathrm m\,50^\mathrm s$ & $+24\deg 10\arcm 57\arcs$ \\
PCCS2 143 G201.43$-$79.27 & ICRF J013738.3$-$243053 & Seyfert 1 Galaxy & $01^\mathrm h\,37^\mathrm m\,41^\mathrm s$ & $-24\deg 31\arcm 08\arcs$ \\
PCCS2 143 G201.45$-$25.27 & PMN J0501$-$0159 & Quasar & $05^\mathrm h\,01^\mathrm m\,18^\mathrm s$ & $-01\deg 58\arcm 49\arcs$ \\
PCCS2 143 G201.84+18.07 & QSO B0735+178 & BL Lac - type object & $07^\mathrm h\,38^\mathrm m\,08^\mathrm s$ & $+17\deg 42\arcm 37\arcs$ \\
PCCS2 143 G203.51$-$24.72 & NGC 1788 & Reflection Nebula & $05^\mathrm h\,06^\mathrm m\,49^\mathrm s$ & $-03\deg 21\arcm 01\arcs$ \\
PCCS2 143 G206.80+35.81 & OJ287 & BL Lac - type object & $08^\mathrm h\,54^\mathrm m\,47^\mathrm s$ & $+20\deg 06\arcm 56\arcs$ \\
PCCS2 143 G207.58$-$23.04 & HH 240/241 & Dark Cloud (nebula) & $05^\mathrm h\,19^\mathrm m\,44^\mathrm s$ & $-05\deg 51\arcm 53\arcs$ \\
PCCS2 143 G208.14+18.76 & QSO B0748+126 & Quasar & $07^\mathrm h\,50^\mathrm m\,51^\mathrm s$ & $+12\deg 31\arcm 45\arcs$ \\
PCCS2 143 G211.31+19.05 & QSO B0754+10 & BL Lac - type object & $07^\mathrm h\,57^\mathrm m\,06^\mathrm s$ & $+09\deg 56\arcm 07\arcs$ \\
PCCS2 143 G215.20$-$24.29 & IC 418 & Planetary Nebula & $05^\mathrm h\,27^\mathrm m\,24^\mathrm s$ & $-12\deg 41\arcm 52\arcs$ \\
PCCS2 143 G216.98+11.39 & QSO B0736+01 & Quasar & $07^\mathrm h\,39^\mathrm m\,20^\mathrm s$ & $+01\deg 37\arcm 24\arcs$ \\
PCCS2 143 G217.68+07.22 & QSO B0723$-$007 & BL Lac - type object & $07^\mathrm h\,25^\mathrm m\,50^\mathrm s$ & $-00\deg 54\arcm 23\arcs$ \\
PCCS2 143 G220.69+24.31 & QSO B0829+047 & BL Lac - type object & $08^\mathrm h\,31^\mathrm m\,45^\mathrm s$ & $+04\deg 29\arcm 22\arcs$ \\
PCCS2 143 G221.44+45.06 & LEDA 1427054(?) & Galactic & $09^\mathrm h\,47^\mathrm m\,59^\mathrm s$ & $+13\deg 16\arcm 51\arcs$ \\
PCCS2 143 G222.60$-$16.18 & QSO B0607$-$157 & Quasar & $06^\mathrm h\,09^\mathrm m\,40^\mathrm s$ & $-15\deg 42\arcm 24\arcs$ \\
PCCS2 143 G223.69$-$34.90 & QSO J0457$-$2324 & Quasar & $04^\mathrm h\,57^\mathrm m\,01^\mathrm s$ & $-23\deg 24\arcm 09\arcs$ \\
PCCS2 143 G227.77+03.13 & QSO J0730$-$116 & Quasar & $07^\mathrm h\,30^\mathrm m\,19^\mathrm s$ & $-11\deg 41\arcm 49\arcs$ \\
PCCS2 143 G228.93+30.92 & 4C 01.24B & Quasar & $09^\mathrm h\,09^\mathrm m\,08^\mathrm s$ & $+01\deg 21\arcm 51\arcs$ \\
PCCS2 143 G229.01$-$36.99 & QSO B0451$-$281 & Quasar & $04^\mathrm h\,53^\mathrm m\,18^\mathrm s$ & $-28\deg 07\arcm 15\arcs$ \\
PCCS2 143 G229.03+13.18 & QSO B0805$-$077 & Quasar & $08^\mathrm h\,08^\mathrm m\,18^\mathrm s$ & $-07\deg 50\arcm 11\arcs$ \\
PCCS2 143 G237.75$-$48.46 & QSO J0403$-$3605 & Quasar & $04^\mathrm h\,03^\mathrm m\,59^\mathrm s$ & $-36\deg 05\arcm 21\arcs$ \\
PCCS2 143 G240.59$-$32.72 & QSO B0521$-$365 & BL Lac - type object & $05^\mathrm h\,22^\mathrm m\,55^\mathrm s$ & $-36\deg 26\arcm 51\arcs$ \\
PCCS2 143 G240.70$-$43.61 & QSO B0426$-$380 & BL Lac - type object & $04^\mathrm h\,28^\mathrm m\,40^\mathrm s$ & $-37\deg 56\arcm 02\arcs$ \\
PCCS2 143 G244.77$-$54.07 & QSO B0332$-$403 & BL Lac - type object & $03^\mathrm h\,34^\mathrm m\,14^\mathrm s$ & $-40\deg 08\arcm 46\arcs$ \\
PCCS2 143 G250.07$-$31.09 & QSO B0537$-$441 & Quasar & $05^\mathrm h\,38^\mathrm m\,46^\mathrm s$ & $-44\deg 04\arcm 58\arcs$ \\
PCCS2 143 G251.50+52.76 & QSO B1055+018 & Quasar & $10^\mathrm h\,58^\mathrm m\,28^\mathrm s$ & $+01\deg 33\arcm 57\arcs$ \\
PCCS2 143 G251.60$-$34.64 & Pictor A & Radio Galaxy & $05^\mathrm h\,19^\mathrm m\,46^\mathrm s$ & $-45\deg 47\arcm 16\arcs$ \\
PCCS2 143 G251.61+21.41 & ICRF J092751.8$-$203451 & Seyfert 1 Galaxy & $09^\mathrm h\,27^\mathrm m\,46^\mathrm s$ & $-20\deg 34\arcm 05\arcs$ \\
PCCS2 143 G251.72$-$35.35 & PKS J0515$-$4556  & Quasar & $05^\mathrm h\,15^\mathrm m\,46^\mathrm s$ & $-45\deg 56\arcm 20\arcs$ \\
PCCS2 143 G251.97$-$38.82 & QSO B0454$-$463 & Quasar & $04^\mathrm h\,55^\mathrm m\,47^\mathrm s$ & $-46\deg 16\arcm 11\arcs$ \\
PCCS2 143 G255.03+81.66 & QSO B1222+216 & Quasar & $12^\mathrm h\,24^\mathrm m\,54^\mathrm s$ & $+21\deg 23\arcm 09\arcs$ \\
PCCS2 143 G261.83$-$60.08 & QSO B0244$-$470 & Quasar & $02^\mathrm h\,46^\mathrm m\,01^\mathrm s$ & $-46\deg 51\arcm 42\arcs$ \\
PCCS2 143 G270.95+24.85 & QSO B1034$-$293 & Quasar & $10^\mathrm h\,37^\mathrm m\,18^\mathrm s$ & $-29\deg 33\arcm 38\arcs$ \\
PCCS2 143 G272.47$-$54.62 & QSO B0252$-$549 & Quasar & $02^\mathrm h\,53^\mathrm m\,27^\mathrm s$ & $-54\deg 40\arcm 55\arcs$ \\
PCCS2 143 G275.27+43.62 & QSO B1127$-$145 & Quasar & $11^\mathrm h\,30^\mathrm m\,05^\mathrm s$ & $-14\deg 50\arcm 22\arcs$ \\
PCCS2 143 G276.08$-$61.76 & QSO B0208$-$512 & BL Lac - type object & $02^\mathrm h\,10^\mathrm m\,50^\mathrm s$ & $-51\deg 01\arcm 16\arcs$ \\
PCCS2 143 G276.72+39.59 & QSO B1124$-$186 & Quasar & $11^\mathrm h\,27^\mathrm m\,05^\mathrm s$ & $-18\deg 56\arcm 45\arcs$ \\
PCCS2 143 G277.16$-$36.04 & LHA 120-N 11 & HII (ionized) region & $04^\mathrm h\,57^\mathrm m\,00^\mathrm s$ & $-66\deg 23\arcm 56\arcs$ \\
PCCS2 143 G277.70$-$32.12 & NGC 2032 & HII (ionized) region & $05^\mathrm h\,35^\mathrm m\,35^\mathrm s$ & $-67\deg 34\arcm 04\arcs$ \\
PCCS2 143 G277.90$-$32.42 & LHA 120-N 57A & HII (ionized) region & $05^\mathrm h\,32^\mathrm m\,18^\mathrm s$ & $-67\deg 41\arcm 55\arcs$ \\
PCCS2 143 G278.38$-$33.32 & RX J0522.0$-$6756 & HII (ionized) region & $05^\mathrm h\,22^\mathrm m\,15^\mathrm s$ & $-67\deg 57\arcm 37\arcs$ \\
PCCS2 143 G279.75$-$34.24 & LHA 120-N 105A & HII (ionized) region & $05^\mathrm h\,09^\mathrm m\,53^\mathrm s$ & $-68\deg 53\arcm 33\arcs$ \\
PCCS2 143 G283.76+74.50 & M87 & Radio Galaxy & $12^\mathrm h\,30^\mathrm m\,49^\mathrm s$ & $+12\deg 24\arcm 08\arcs$ \\
PCCS2 143 G284.15+14.21 & QSO B1104$-$445 & Quasar & $11^\mathrm h\,07^\mathrm m\,04^\mathrm s$ & $-44\deg 49\arcm 16\arcs$ \\
PCCS2 143 G284.85+66.07 & 4C 04.42 & Quasar & $12^\mathrm h\,22^\mathrm m\,26^\mathrm s$ & $+04\deg 13\arcm 31\arcs$ \\
PCCS2 143 G286.37$-$27.15 & QSO J0635$-$7516 & Quasar & $06^\mathrm h\,35^\mathrm m\,45^\mathrm s$ & $-75\deg 16\arcm 23\arcs$ \\
PCCS2 143 G289.22+22.93 & QSO B1144$-$3755 & BL Lac - type object & $11^\mathrm h\,46^\mathrm m\,57^\mathrm s$ & $-38\deg 12\arcm 43\arcs$ \\
PCCS2 143 G289.94+64.36 & 3C 273 & Quasar & $12^\mathrm h\,29^\mathrm m\,06^\mathrm s$ & $+02\deg 03\arcm 12\arcs$ \\
PCCS2 143 G293.83$-$31.37 & QSO B0454$-$810 & Quasar & $04^\mathrm h\,50^\mathrm m\,10^\mathrm s$ & $-80\deg 59\arcm 58\arcs$ \\
PCCS2 143 G296.96$-$05.76 & QSO B1145$-$676 & Quasar & $11^\mathrm h\,47^\mathrm m\,35^\mathrm s$ & $-67\deg 53\arcm 21\arcs$ \\
PCCS2 143 G297.31$-$07.74 & ICRF J114553.6$-$695401 & Seyfert 1 Galaxy & $11^\mathrm h\,45^\mathrm m\,49^\mathrm s$ & $-69\deg 53\arcm 58\arcs$ \\
PCCS2 143 G298.01$-$18.28 & QSO B1057$-$797 & BL Lac - type object & $10^\mathrm h\,58^\mathrm m\,48^\mathrm s$ & $-80\deg 04\arcm 05\arcs$ \\
PCCS2 143 G301.62+37.06 & QSO J1246$-$2547 & Quasar & $12^\mathrm h\,46^\mathrm m\,48^\mathrm s$ & $-25\deg 47\arcm 25\arcs$ \\
PCCS2 143 G304.15$-$72.18 & 2MASX J00491665-4457110 & Quasar & $00^\mathrm h\,49^\mathrm m\,19^\mathrm s$ & $-44\deg 56\arcm 11\arcs$ \\
PCCS2 143 G305.09+57.06 & 3C 279 & Quasar & $12^\mathrm h\,56^\mathrm m\,09^\mathrm s$ & $-05\deg 47\arcm 30\arcs$ \\
PCCS2 143 G308.33+09.56 & VLBI 1323$-$527 & Quasar & $13^\mathrm h\,26^\mathrm m\,54^\mathrm s$ & $-52\deg 56\arcm 09\arcs$ \\
PCCS2 143 G309.51+19.42 & Centaurus A & Radio Galaxy & $13^\mathrm h\,25^\mathrm m\,27^\mathrm s$ & $-43\deg 00\arcm 28\arcs$ \\
PCCS2 143 G313.43$-$18.84 & QSO B1610$-$771 & Quasar & $16^\mathrm h\,17^\mathrm m\,43^\mathrm s$ & $-77\deg 16\arcm 58\arcs$ \\
PCCS2 143 G315.80$-$36.52 & QSO J2147$-$7536 & Quasar & $21^\mathrm h\,47^\mathrm m\,07^\mathrm s$ & $-75\deg 36\arcm 19\arcs$ \\
PCCS2 143 G316.49+21.14 & TGU H1970 P1 & Dark Cloud (nebula) & $13^\mathrm h\,57^\mathrm m\,45^\mathrm s$ & $-39\deg 58\arcm 26\arcs$ \\
PCCS2 143 G320.02+48.37 & QSO B1334$-$127 & Quasar & $13^\mathrm h\,37^\mathrm m\,40^\mathrm s$ & $-12\deg 57\arcm 24\arcs$ \\
PCCS2 143 G320.18$-$62.12 & QSO B2355$-$534 & Quasar & $23^\mathrm h\,57^\mathrm m\,56^\mathrm s$ & $-53\deg 10\arcm 22\arcs$ \\
PCCS2 143 G321.30$-$40.64 & ESO 75-41 & Seyfert 1 Galaxy & $21^\mathrm h\,57^\mathrm m\,11^\mathrm s$ & $-69\deg 41\arcm 43\arcs$ \\
PCCS2 143 G321.36+56.24 & QSO J1332$-$0509 & Quasar & $13^\mathrm h\,32^\mathrm m\,04^\mathrm s$ & $-05\deg 09\arcm 41\arcs$ \\
PCCS2 143 G321.43+17.26 & QSO B1424$-$41 & Quasar & $14^\mathrm h\,27^\mathrm m\,53^\mathrm s$ & $-42\deg 06\arcm 34\arcs$ \\
PCCS2 143 G325.18+25.59 & ICRF J142741.3$-$330531 & BL Lac - type object & $14^\mathrm h\,27^\mathrm m\,40^\mathrm s$ & $-33\deg 05\arcm 15\arcs$ \\
PCCS2 143 G326.97$-$05.90 & 3FGL J1617.4$-$5846 & Quasar & $16^\mathrm h\,17^\mathrm m\,18^\mathrm s$ & $-58\deg 47\arcm 57\arcs$ \\
PCCS2 143 G328.13$-$12.43 & ICRF J170336.5$-$621240 & BL Lac - type object & $17^\mathrm h\,03^\mathrm m\,44^\mathrm s$ & $-62\deg 12\arcm 08\arcs$ \\
PCCS2 143 G328.22+18.96 & ICRF J145427.4$-$374733 & Seyfert 1 Galaxy & $14^\mathrm h\,54^\mathrm m\,25^\mathrm s$ & $-37\deg 47\arcm 40\arcs$ \\
PCCS2 143 G335.72$-$64.05 & QSO B2326$-$477 & Quasar & $23^\mathrm h\,29^\mathrm m\,17^\mathrm s$ & $-47\deg 30\arcm 19\arcs$ \\
PCCS2 143 G340.68+27.57 & QSO B1514$-$24 & BL Lac - type object & $15^\mathrm h\,17^\mathrm m\,43^\mathrm s$ & $-24\deg 22\arcm 18\arcs$ \\
PCCS2 143 G347.47$-$12.47 & ICRF J180957.8$-$455241 & BL Lac - type object & $18^\mathrm h\,10^\mathrm m\,01^\mathrm s$ & $-45\deg 52\arcm 44\arcs$ \\
PCCS2 143 G351.27+40.12 & QSO J1512$-$0906 & Quasar & $15^\mathrm h\,12^\mathrm m\,51^\mathrm s$ & $-09\deg 07\arcm 01\arcs$ \\
PCCS2 143 G352.45$-$08.41 & ICRF J180242.6$-$394007 & Quasar & $18^\mathrm h\,02^\mathrm m\,39^\mathrm s$ & $-39\deg 39\arcm 39\arcs$ \\
PCCS2 143 G352.59$-$40.37 & QSO B2052$-$474 & Quasar & $20^\mathrm h\,56^\mathrm m\,12^\mathrm s$ & $-47\deg 14\arcm 31\arcs$ \\

  \end{longtable}
}

\subsection{NPIPE processing}

The \Planck\ NPIPE data release (also known as PR4) differs from earlier data releases in several critical ways \citep{planck2020-LVII}.  
It is the first data release based on a single, stand-alone pipeline that processes both raw LFI and HFI data into calibrated, bandpass-corrected, noise-subtracted frequency maps. 
The release comprises frequency maps, Monte Carlo simulations, and calibrated timelines which we used as inputs to our processing.  The timelines are accompanied by an updated instrument model that includes notable corrections to HFI polarization angles and efficiencies.

The NPIPE release is the first to include detector data acquired during so-called repointing maneuvers.  Including these 4-minute periods between stable science scans not only increases integration time by about 9\% but also improves sky sampling by adding detector samples between the regular scanning paths, which are separated by two arcminutes.  NPIPE suppresses degree-scale noise in the maps by modeling the $1/f$ noise fluctuations in the data with $167\,$ms noise offsets.  In contrast, earlier LFI data were fitted with one-second offsets and earlier HFI data with full pointing period ($45$-minute) offsets.

We list here other systematic effects and artifacts corrected in NPIPE processing:
\begin{itemize}
    \item electrical interference from housekeeping electronics (for LFI) and the 4-kelvin cooler (for HFI);
    \item jumps in signal offset;
    \item cosmic ray glitches (HFI);
    \item bolometric nonlinearity (HFI);
    \item thermal common modes (HFI);
    \item ADC nonlinearity (HFI);
    \item gain fluctuations;
    \item bandpass mismatch from both continuum emission and CO lines;
    \item bolometric transfer function and transfer function residuals (HFI);
    \item far-sidelobe pickup.
\end{itemize}
In addition, NPIPE removes time-dependent signals from the orbital dipole and Zodiacal light that would otherwise compromise mapmaking.

\section{Technical approach}
\label{sec:method}

As noted above, \Planck\ sees most compact sources on the sky only once during each six-month sky survey.  We call the roughly week-long period during which the \Planck\ detectors collect time samples in the vicinity of a source an \emph{observation}.  During each observation, the orientation of the detectors with respect to the source remains fixed.  Consequently, individual detectors cannot resolve the polarized flux density from the intensity in a single observation. To overcome this restriction, different detectors in a given \Planck\ band are given different polarization angles (see Fig.~\ref{fig:my_label}).

\subsection{Individual detectors}

To solve for the polarized flux density, we first measured the polarization-modulated flux density seen by each of the \Planck\ detectors.  The data model can be cast in the typical linear regression form:
\begin{equation}
    \vec{d} = F\vec{a} + \vec{n},
\end{equation}
where
\begin{itemize}
    \item $\vec{d}$ is the vector of detector samples that fall within some fitting radius from the source coordinates;
    \item $F$ contains the time-domain templates representing the source and the background. The primary template in $F$ is the detector point spread function (PSF).  This is the beam pattern produced when a detector with finite resolution scans a point source.
    \item $\vec{a}$ contains the template amplitudes;
    \item $\vec{n}$ is the detector noise.
\end{itemize}
The maximum likelihood solution for the template amplitudes, $\vec{a}$, is:
\begin{equation} \label{eq:amplitudes}
    \vec a = \left(F^\mathrm{T} N^{-1} F\right)^{-1} F^\mathrm{T} N^{-1} \vec d,
\end{equation}
where $N$ is the noise covariance matrix $N = \langle \vec {n}\vec{n}^\mathrm{T} \rangle$.  The \Planck\ PR4 timelines used as inputs are calibrated and destriped, making a diagonal noise matrix, $N = \sigma^2 \mathbf{1}$, an excellent approximation. Furthermore, the detector samples in $\vec{d}$ are separated in time by 60\,s, the time it takes for \Planck\ to spin around, effectively breaking sample-sample noise correlations.

The PSF must be rotated into the appropriate orientation and centered at the cataloged source position.  The fitted amplitude of the PSF template is readily translated into a flux density across the detector pass band.  Since the detector data are calibrated in thermodynamic units, the fitted amplitudes must be translated into flux density using the unit conversion factors derived from ground-measured band-passes. The full unit conversion is 
\begin{equation}
    f = T_\mathrm{max}\,\Omega_\mathrm{det}\,u_\mathrm{det},
\end{equation}
where $T_\mathrm{max}$ is the peak temperature read off the PSF fit, $\Omega_\mathrm{det}$ is the solid angle of the PSF and $u_\mathrm{det}$ is the unit conversion factor from kelvin to MJy/sr, derived by convolving the detector bandpass with a standard source spectrum (slope $\alpha=-1$).

Additional columns in $F$ model the background [foreground] emission.  There is always a constant offset template, but we also considered higher-order templates such as slopes.  Keeping the fitting radius close to the beamwidth limits the number of modes needed to fully model the background [foreground].

Small offset between the catalog source position and apparent source location can be modeled by adding spatial derivatives of the PSF in $F$.  Such offsets naturally rise if the source position was originally fitted over a pixelized sky map.  There can also be small pointing errors in the \Planck\ data themselves.  Finally, the center of the PSF template is slightly arbitrary and can be defined as either the peak or the center of weight.  For vast majority of our targets, including position corrections in the fit increases uncertainty without a significant improvement in the quality of the fit.  The few cases where the improvement is statistically significant (source is very bright), the change in measured flux density is less than a percent.  Therefore the reported flux densities are based on fits directly at the catalog positions.

\subsection{Polarized flux density}

For a given observation, each detector produces one data point: the polarization-modulated flux density:
\begin{equation} \label{eq:signal_model1}
    f_d = I + \eta_d \left[Q \cos(2(\psi+\psi_d)) + U \sin(2(\psi+\psi_d))\right],
\end{equation}
where
\begin{itemize}
    \item $f_d$ is the polarization-modulated flux density measured with detector $d$;
    \item $I$, $Q$ and $U$ are the Stokes parameters of the source flux density (CMB detectors are not typically sensitive to $V$, the circular polarization component);
    \item $\eta$ is the detector polarization efficiency -- an ideal polarization-sensitive detector would have $\eta=1$;
    \item $\psi$ represents the angle between the source coordinate system and the focal-plane coordinate system;
    \item $\psi_d$ is the polarization sensitive direction of the detector in the focal-plane coordinates.
\end{itemize}
A vector of detector fluxes, $\vec f$, can be modeled as
\begin{equation}
    \vec f = P \vec m + \vec n,
\end{equation}
where each row of the pointing matrix, $P$, comprises the pointing weights, $[1, \eta\cos(2(\psi+\psi_d)), \eta\sin(2(\psi+\psi_d))]$, for one detector, and the ``map'' vector contains the three Stokes parameters: $m=[I,Q,U]^\mathrm T$.  The noise vector, $n$, contains the measurement uncertainty for each detector.  Analogous to Eq.~(\ref{eq:amplitudes}), the maximum likelihood solution reads:
\begin{equation}
    \vec m = \left(P^\mathrm T N^{-1} P\right)^{-1} P^\mathrm{T} N^{-1} \vec f.
\end{equation}
The theory of linear regression also provides the covariances:
\begin{equation}
    N_m = \langle \Delta \vec{m} \Delta \vec{m}^\mathrm{T} \rangle
    = \left(P^\mathrm T N^{-1} P\right)^{-1}
    = \begin{bmatrix}
    \sigma_I^2 & \sigma_{IQ} & \sigma_{IU} \\
     & \sigma_Q^2 & \sigma_{QU} \\
     && \sigma_U^2
    \end{bmatrix},
\end{equation}
where we have omitted the redundant elements in the symmetric covariance matrix.

In order to derive the polarization fraction and polarization angle, we defined two estimators:
\begin{equation} \label{eq:biased_estimators}
    \hat{p} = \frac{\sqrt{Q^2 + U^2}}{I}
    \quad\mathrm{and}\quad
    \hat{\psi} = \frac{1}{2}\arctan{\frac{U}{Q}}.
\end{equation}
While these estimators are textbook definitions of the polarization parameters, they are biased by uncertainties in $I$, $Q$, and $U$.  Finding unbiased or minimally biased estimators for these parameters turns out to be complicated and depends on the signal-to-noise ratio (S/N) of the $I,Q,U$ estimates.  For the modest S/N in our case, we have adopted the asymptotic estimator from \cite{Montier_2015}:
\begin{equation}
    \hat{p}_\mathrm{AS} = \begin{cases}
    \sqrt{\hat{p}^2 - b^2} & \mathrm{if} \,\, \hat p > b, \\
    0 & \mathrm{otherwise}
    \end{cases},
\end{equation}
where
\begin{eqnarray}
    b^2 & = & \frac{
            \widetilde\sigma_U^2\cos^2\left(2\psi_0-\theta\right)
            + \widetilde\sigma_Q^2\sin^2\left(2\psi_0-\theta\right)
        }{I_0^2} \\
    \theta & = &
        \frac{1}{2}\arctan\left(
            \frac{2\rho\sigma_Q\sigma_U}{\sigma_Q^2 - \sigma_U^2}
        \right) \\
    \widetilde\sigma_Q^2 & = &
        \sigma_Q^2\cos^2\theta + \sigma_U^2\sin^2\theta + \rho\sigma_Q\sigma_U\sin 2\theta \\
    \widetilde\sigma_U^2 & = &
        \sigma_Q^2\sin^2\theta + \sigma_U^2\cos^2\theta - \rho\sigma_Q\sigma_U\sin 2\theta \\
    \rho & = & \frac{\sigma_{QU}}{\sigma_Q\sigma_U}.
\end{eqnarray}
The estimator from Eq.~(\ref{eq:biased_estimators}) is typically substituted in place of the true polarization angle, $\psi_0$. 

\begin{figure*}
    \centering
    \includegraphics[trim=0 0 0 0,clip,width=\textwidth]{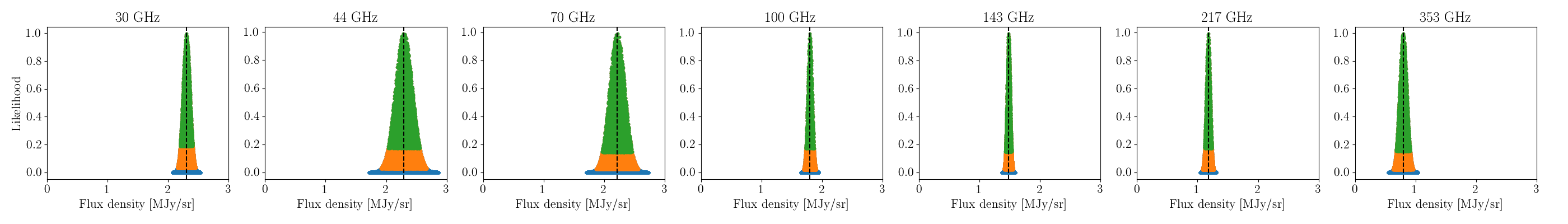}
    \includegraphics[trim=0 0 0 0,clip,width=\textwidth]{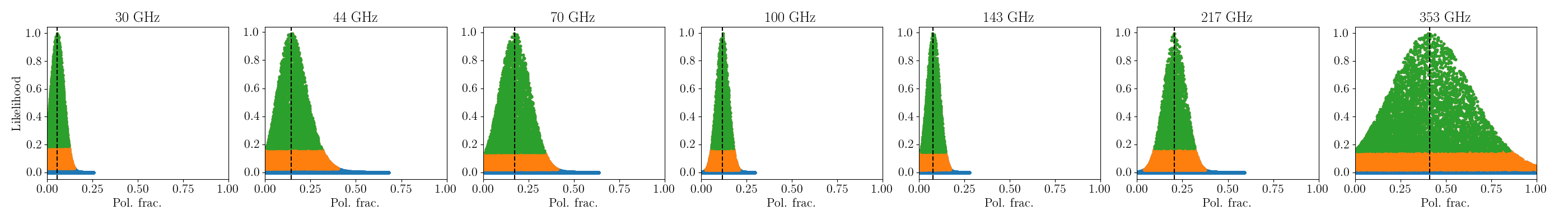}
    \includegraphics[trim=0 0 0 0,clip,width=\textwidth]{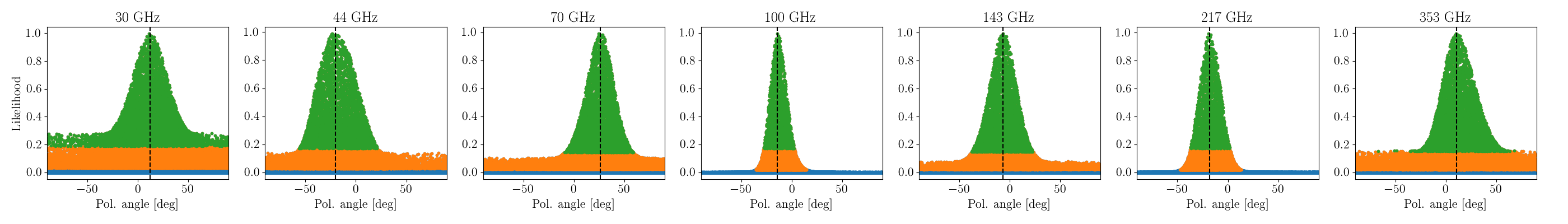}
    \includegraphics[trim=0 0 0 0,clip,width=\textwidth]{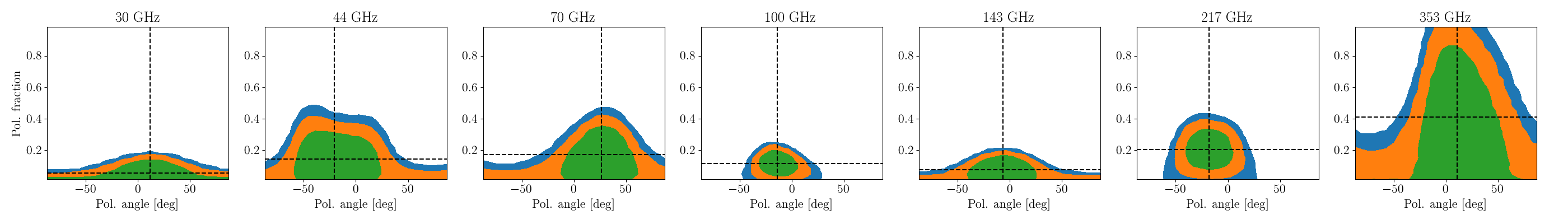}
    \caption{
        100,000 sampled values of the likelihood function, projected to the intensity axis (top row), polarization fraction axis (2nd row) or polarization angle axis (3rd row).  The source, PCCS2\,143\,G052.40$-$36.49 also known as 4C\,$-$02.81, was chosen because it has both positive detections of and upper limits on polarization fraction.  The likelihood threshold that includes 68\,\% of the total likelihood is indicated in green and 95\,\% in orange.  Samples with likelihood lower than 95\,\% of the distribution are in blue.  When the green part of the polarization fraction distribution excludes zero, we report a 68\,\% confidence interval.  When zero is included, we report a 95\,\% upper limit. The bottom row presents the maximum of the likelihood function in the polarization fraction-polarization angle plane.  The dashed lines indicate the maximum likelihood values.  The likelihoods are based on the first survey shown in Fig.~\ref{fig:pol_frac_uncertainty_source}.
    }
    \label{fig:pol_frac_uncertainty}
\end{figure*}

Confidence regions for the polarization parameters are determined using a Monte Carlo method: we sampled 100,000 triplets of true $(I_0, p_0, \psi_0)$ and recorded the likelihood of the measured $(I, p, \psi)$. From those likelihoods, we inferred the 68\% and 95\% confidence regions for the true polarization parameters. The Bayesian likelihood, $\mathcal L(I, p, \psi|I_0, p_0, \psi_0)$, is defined in \cite{Montier_2015}, Eqs.~(3) and (23).  Examples of the sampled likelihood, projected to the $p$-axis are shown in Fig.~\ref{fig:pol_frac_uncertainty}.

\begin{figure*}
    \centering
    \includegraphics[trim=45 30 0 60,clip,width=\textwidth]{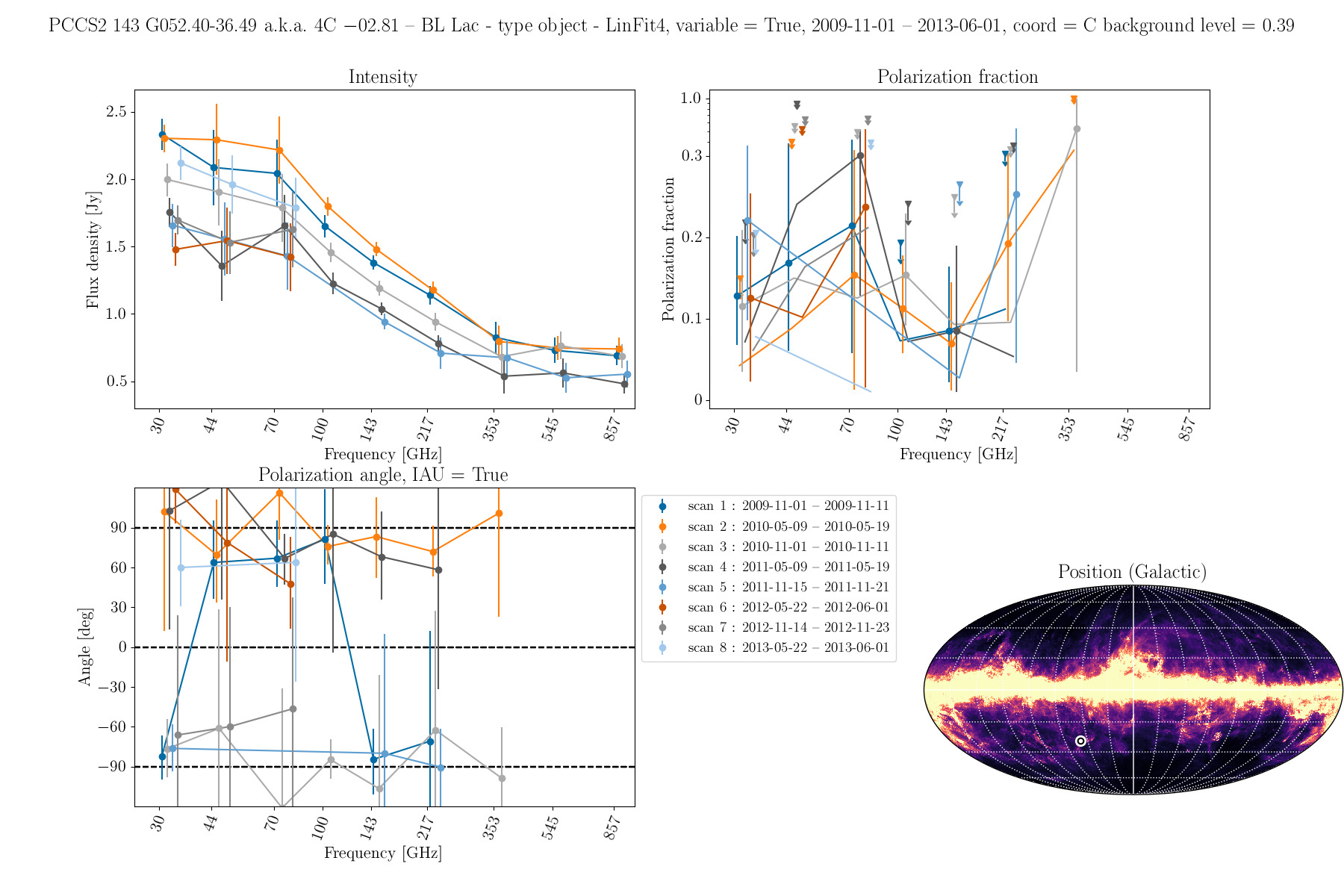}
    \caption{Example of a full diagnostic plot showing intensity, polarization fraction, polarization angle, and position for PCCS2 143  G052.40-36.49.  The polarization fraction axis is nonlinear between $0.3$ and $1.0$ to accommodate all values. The lines and points of different colours mark different observing epochs, separated by 6 months. The Galactic-coordinate map (HEALPix formatted, \citealt{2005ApJ...622..759G}) in the lower right panel shows the position of the source on top of the Planck 857\,GHz intensity map. Plots and maps similar to this are provided for each source and each number of jointly fit frequencies.}
    \label{fig:pol_frac_uncertainty_source}
\end{figure*}

\subsection{Multifrequency fits}

In addition to tabulating flux densities at each \Planck\ frequency, we provide multifrequency fits in order to improve the S/N of the measured polarized flux densities.  A fit such as this necessarily requires some accommodation of the source spectral energy distribution (SED).  We extended the signal model in Eq.~(\ref{eq:signal_model1}) to:
\begin{equation} \label{eq:signal_model2}
    f_d = \left(
        \frac{\nu_d}{\nu_0}\right)^\alpha
    \Big\{
    I + \eta_d \left[Q \cos(2(\psi+\psi_d)) + U \sin(2(\psi+\psi_d))\right]
    \Big\},
\end{equation}
where $\alpha$ is the spectral index of the source, $\nu_d$ is the central frequency of detector $d$, and $\nu_0$ is the noise-weighted central frequency of all detectors that are being combined.  This model implies strong constraints on the fit:
\begin{itemize}
    \item the SED is approximated as a power-law across the fitted frequencies;
    \item the spectral indices of intensity and polarized emission are the same;
    \item the polarization fraction and angle do not depend on the frequency.
\end{itemize}
These approximations obviously break down as increasingly wider frequency ranges are considered.  Hence we provide multi-frequency fits only up to three adjacent frequency bands.  Within these bands, and to the extent the approximations are valid, the additional constraints greatly improve the S/N of our fits.

\subsection{Software}

Our flux-fitting software is implemented in Python-3 and released as open source software.\footnote{\url{https://github.com/hpc4cmb/tod2flux}} The package includes \Planck-specific specializations but they are modularized and the software can be adapted for other experimental data sets that can be sampled into the small data set format used as inputs to the calculation.

\section{The catalog}
\label{sec:catalog}

Our catalog of time- and frequency-dependent, polarized, flux densities is spread over multiple comma-separated value (CSV) files.  Each file features a single source and the entries correspond to a fixed number (ranging from 1 to 3) of frequency bands averaged together.  The filenames are of the form \texttt{results\_<PCCSNAME>\_nfreq=<NFREQ>.csv}, where \texttt{<PCCSNAME>} is the name of the source in the 143\,GHz \Planck\ PCCS2 catalog and \texttt{<NFREQ>} is the number of frequency bands averaged for each entry (1, 2, or 3). The fields of the files are
\begin{enumerate}
    \item \texttt{band(s)} -- nominal \Planck\ bands comprising the entry.
    \item \texttt{freq [GHz]} -- noise-weighted average of the nominal frequencies.
    \item \texttt{start} -- approximate UT calendar date of the start of the observations used in deriving the entry.
    \item \texttt{stop} --  approximate UT calendar date of the end of the observations.
    \item \texttt{I flux [mJy]} -- intensity flux density at the effective central frequency, not color-corrected to match \texttt{freq [GHz]}.
    \item \texttt{I error [mJy]} -- 1-$\sigma$ uncertainty of the \texttt{I flux [mJy]}
    \item \texttt{Q flux [mJy]} -- linear $Q$-polarization flux density in IAU convention and Equatorial (J2000) coordinates.
    \item \texttt{Q error [mJy]} -- 1-$\sigma$ uncertainty of \texttt{Q flux [mJy]}.
    \item \texttt{U flux [mJy]} -- linear $U$-polarization flux density in IAU convention and Equatorial (J2000) coordinates.
    \item \texttt{U error [mJy]} -- 1-$\sigma$ uncertainty of \texttt{U flux [mJy]}.
    \item \texttt{Pol.Frac} -- debiased polarization fraction in the range [$0\dots1$].
    \item \texttt{PF low lim} -- Bayesian 16\% quantile lower limit of \texttt{Pol.Frac} or zero.
    \item \texttt{PF high lim} -- If \texttt{PF low lim} is nonzero the Bayesian 16\% quantile upper limit of \texttt{Pol.Frac}.  Otherwise the 95\% upper limit.
    \item \texttt{Pol.Ang [deg]} -- Polarization angle in IAU convention and Equatorial (J2000) coordinates.
    \item \texttt{PA low lim} -- Bayesian 16\% quantile lower limit of \texttt{Pol.Ang [deg]}.
    \item \texttt{PA high lim} -- Bayesian 16\% quantile upper limit of \texttt{Pol.Ang [deg]}.
\end{enumerate}

\begin{figure*}
    \centering
    \includegraphics[trim=30 40 60 80,clip,width=\textwidth]{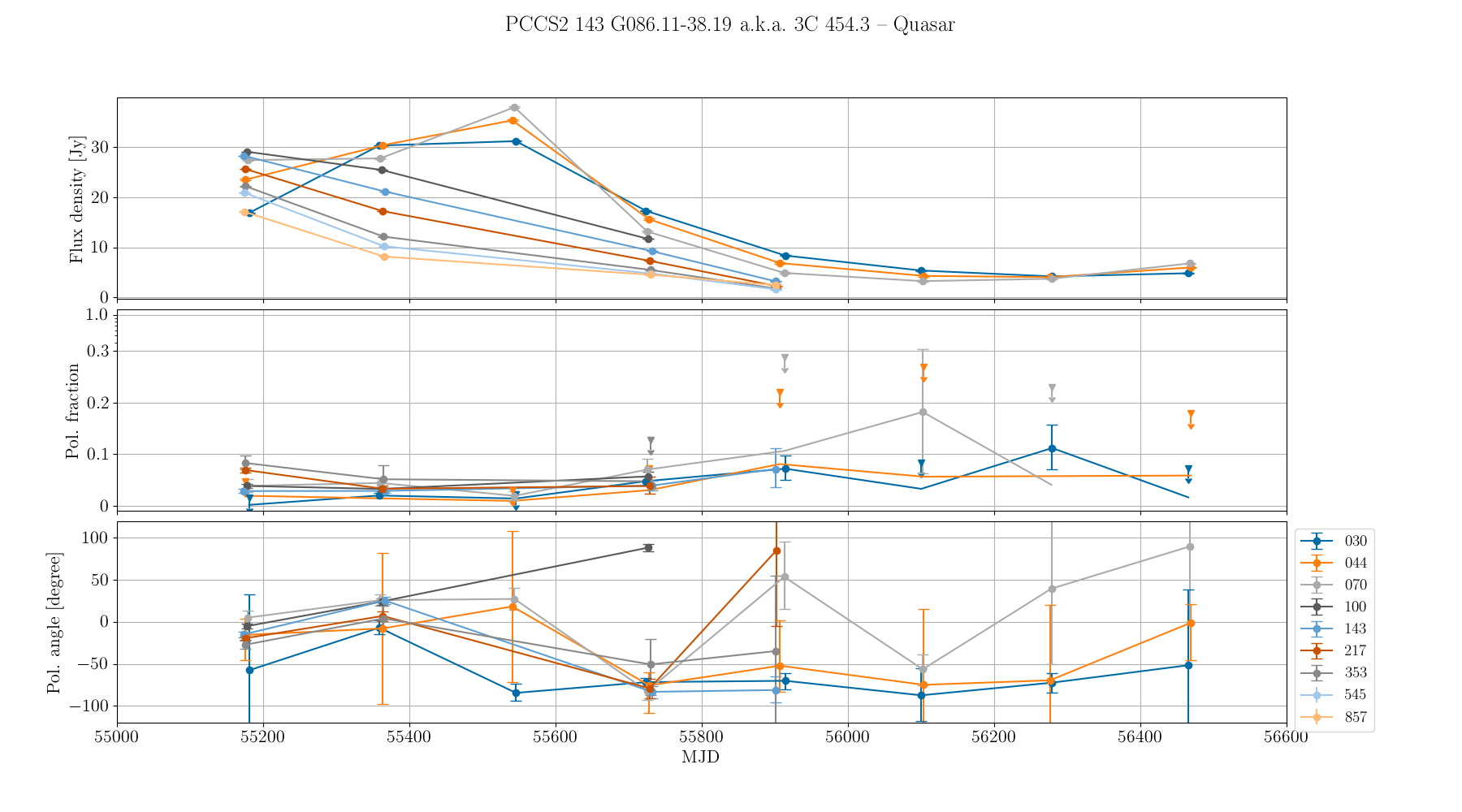}
    \caption{
        Time-series plot of our flux density fits to 3C\,454.3 observations.  The HFI data for the third pass are missing due to quality flagging.
    }
    \label{fig:3c454_vs_time}
\end{figure*}

An excerpt of one catalog file is shown in Appendix~\ref{app:sample}.  The CSV file is accompanied by two diagnostic plots that visualize its contents:
\begin{itemize}
    \item \texttt{flux\_fit\_<PCCSNAME>\_nfreq=<NFREQ>.png}
    \item \texttt{results\_<PCCSNAME>\_nfreq=<NFREQ>.replot.png}
\end{itemize}
The first of these shows the fit data as a function of frequency (e.g., Fig.~\ref{fig:pol_frac_uncertainty_source}) and the second as a function of time (e.g., Fig.~\ref{fig:3c454_vs_time}).

For convenience, we also provide a single catalog file, \texttt{combined.csv}, that combines entries from all individual \texttt{results\_*csv} files so that it includes data for all 153 sources.  This file has three additional columns appended to the beginning of each row:
\begin{itemize}
    \item \texttt{target} -- Name of the target in the \Planck\ 143GHz PCCS2 catalog;
    \item \texttt{RA} -- Right ascension in degrees (J2000);
    \item \texttt{Dec} -- Declination in degrees (J2000).
\end{itemize}

The flux densities in our catalog are not color-corrected.  That is, we made no effort to translate the measured flux density to the nominal observing frequency.  The reason is that the spectral indices of the sources can vary as a function of time and across the \Planck\ pass-band, and the small corrections (of order a few percent) depend on a source's spectral index.  Instead,  we present in Table~\ref{tab:cfreq} the effective central frequencies of the \Planck\ bands for a number of spectral slopes.  The formulas for evaluating the central frequency are laid out in \cite{planck2013-p03d}.

Our catalog, PCCS-PV, will be available at the ESA Planck Legacy Archive.\footnote{\url{https://www.cosmos.esa.int/web/planck/pla}}  and at the NASA/IPAC Infrared Science Archive. \footnote{\url{https://irsa.ipac.caltech.edu/Missions/planck.html}}

\begin{table*}[ht!]
  \renewcommand{\arraystretch}{1.2}
  \renewcommand{\tabcolsep}{0.3cm}
  \begin{center}
    \caption{Effective central frequency in GHz as a function of the nominal frequency and spectral index across the pass-band.}
    \label{tab:cfreq}
    \begin{tabular}{r@{}l@{\,\,\dots\,\dots\,\dots\qquad} r r r r r r r}
    \hline\hline
      \multicolumn{2}{c}{Nominal frequency}& \multicolumn{7}{c}{Spectral index $\alpha$}\\\cline{3-9}
      \multicolumn{2}{c}{(GHz)}&
      \multicolumn{1}{c}{$-2$}&
      \multicolumn{1}{c}{$-1$}&
      \multicolumn{1}{c}{$0$}&
      \multicolumn{1}{c}{$1$}&
      \multicolumn{1}{c}{$2$}&
      \multicolumn{1}{c}{$3$}&
      \multicolumn{1}{c}{$4$}\\
      \hline
      30 &   &   28.10 &   28.28 &   28.46 &   28.64 &   28.83 &   29.03 &   29.22 \\
      44 &   &   43.79 &   43.94 &   44.09 &   44.25 &   44.40 &   44.55 &   44.69 \\
      70 &   &   69.43 &   69.90 &   70.36 &   70.80 &   71.24 &   71.66 &   72.07 \\
      100&   &   99.36 &  100.28 &  101.23 &  102.22 &  103.28 &  104.46 &  105.95 \\
      143&   &  139.82 &  141.17 &  142.57 &  144.04 &  145.59 &  147.28 &  149.18 \\
      143& P &  139.29 &  140.63 &  142.04 &  143.52 &  145.10 &  146.83 &  148.77 \\
      217&   &  218.08 &  220.03 &  222.02 &  224.05 &  226.14 &  228.30 &  230.60 \\
      217& P &  217.18 &  219.18 &  221.23 &  223.32 &  225.47 &  227.69 &  230.04 \\
      353&   &  355.38 &  358.07 &  360.82 &  363.62 &  366.47 &  369.40 &  372.44 \\
      353& P &  354.57 &  357.22 &  359.93 &  362.70 &  365.53 &  368.45 &  371.50 \\
      545&   &  547.42 &  552.80 &  558.00 &  563.01 &  567.84 &  572.45 &  576.83 \\
      857&   &  845.64 &  853.99 &  862.02 &  869.76 &  877.23 &  884.40 &  891.26 \\
      \hline
    \end{tabular}
    \tablefoot{  Frequencies with ``P'' suffix denote the subsets of detectors that are polarization-sensitive. Spectral index is defined by $S_\nu \propto \nu^\alpha$.}
  \end{center}
\end{table*}

\section{Results and discussion}
\label{sec:results}

\subsection{Polarized signal-to-noise ratio}

\begin{figure*}
    \centering
    \includegraphics[trim=100 50 120 100,clip,width=\textwidth]{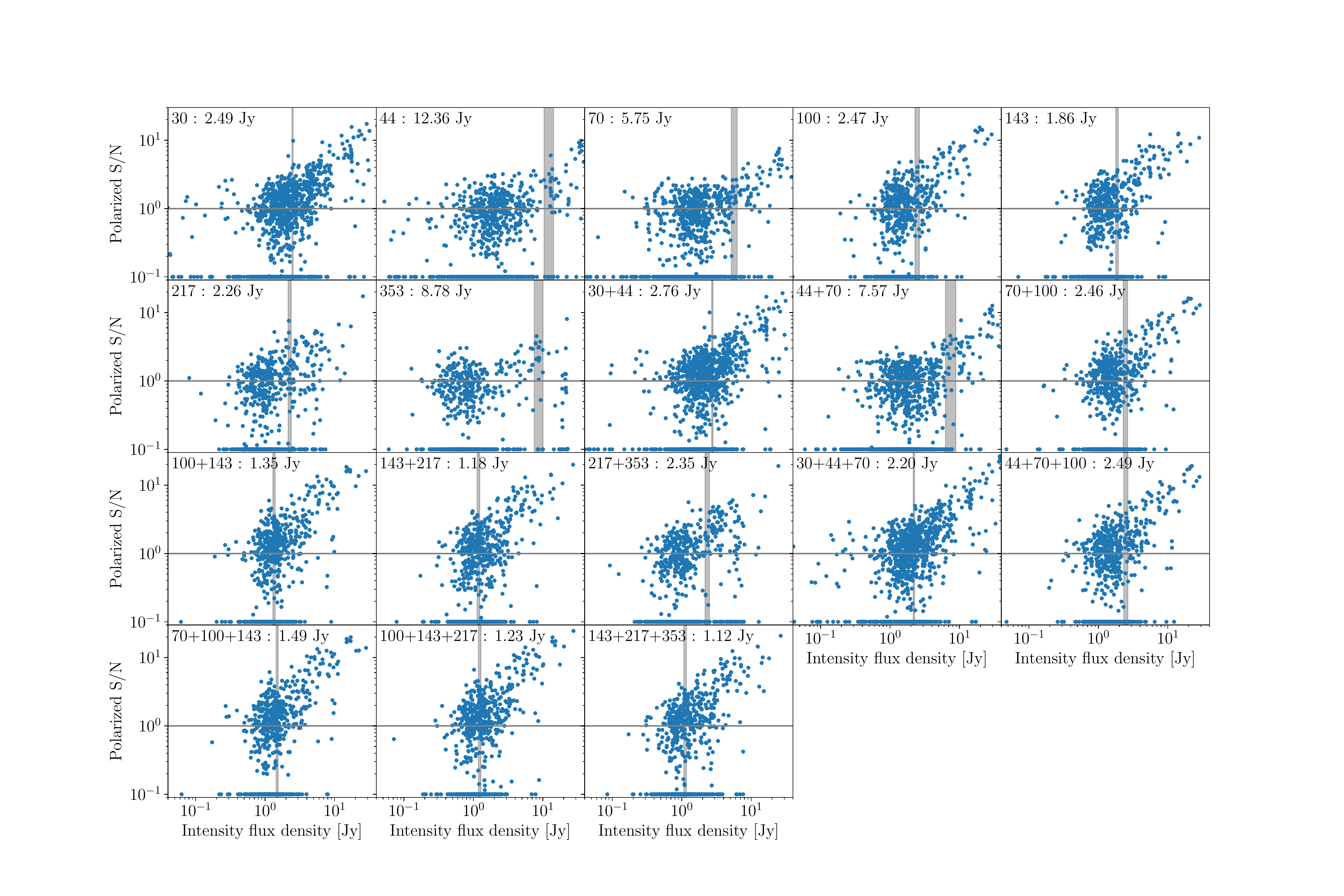}
    \caption{
        Polarized S/N:  for each polarization-sensitive observation in our catalog, we plot the de-biased polarized flux density S/N against the intensity flux density.  Observing bands are indicated in the labels.  For cases where de-biasing renders a S/N less than 0.1, we plot a data point at 0.1.  The vertical grey bands indicate the lowest and narrowest flux density bin that has a median S/N of 1.0 and at least 20 entries.  The center of the bin is spelled out in the label.
    }
    \label{fig:polarized_snr}
\end{figure*}

The targets in our catalog vary in intensity, polarization fraction and spectral properties.  We explore the polarized S/N of our catalog in Fig.~\ref{fig:polarized_snr}.  The Figure shows the scatter in polarized S/N as a function of the intensity flux density and identifies a threshold beyond which half of the observations reach S/N of one.  The threshold varies considerably across the \Planck\ frequencies and obviously depends on the spectral and polarization properties of our catalog.

Our most sensitive, single frequency measurements of polarized flux density come from the 143\,GHz band where the S/N is higher than unity in a majority of observations when the source is brighter than 1.86\,Jy.  Fitting for the source spectral index and combining frequencies gives us a combined threshold better than 1.5\,Jy in all frequency combinations involving 143\,GHz.  Combining the \Planck\ LFI frequencies (30--70\,GHz) has a combined threshold of 2.20\,Jy.

\subsection{Characterization}\label{s:char}

We made use of both spectral information and classifications from the literature, mainly NED \citep{1995ASSL..203...95H} and SIMBAD \citep{2000A&AS..143....9W}, to divide our sample of 153 sources into several categories.  The majority are extragalactic, but nine (6\%) are Galactic sources such as \ion{H}{ii} regions or planetary nebulae, and five more are \ion{H}{ii} regions located in the Magellanic Clouds.  Many of the extragalactic objects are well-studied radio sources such as 3C\,84 or M87.  We characterize the 139 extragalactic sources as $\sim$25\% BL Lac objects, $\sim$60\% quasars and $\sim$10\% Seyfert-1 galaxies.  Some sources exhibit both synchrotron emission with a falling spectrum and dust emission at the highest \Planck\ frequencies (see Fig.~\ref{fig:nonvarsources} for an example).

\begin{figure}
    \centering
    \includegraphics[trim=70 415 670 60,clip,width=\columnwidth]{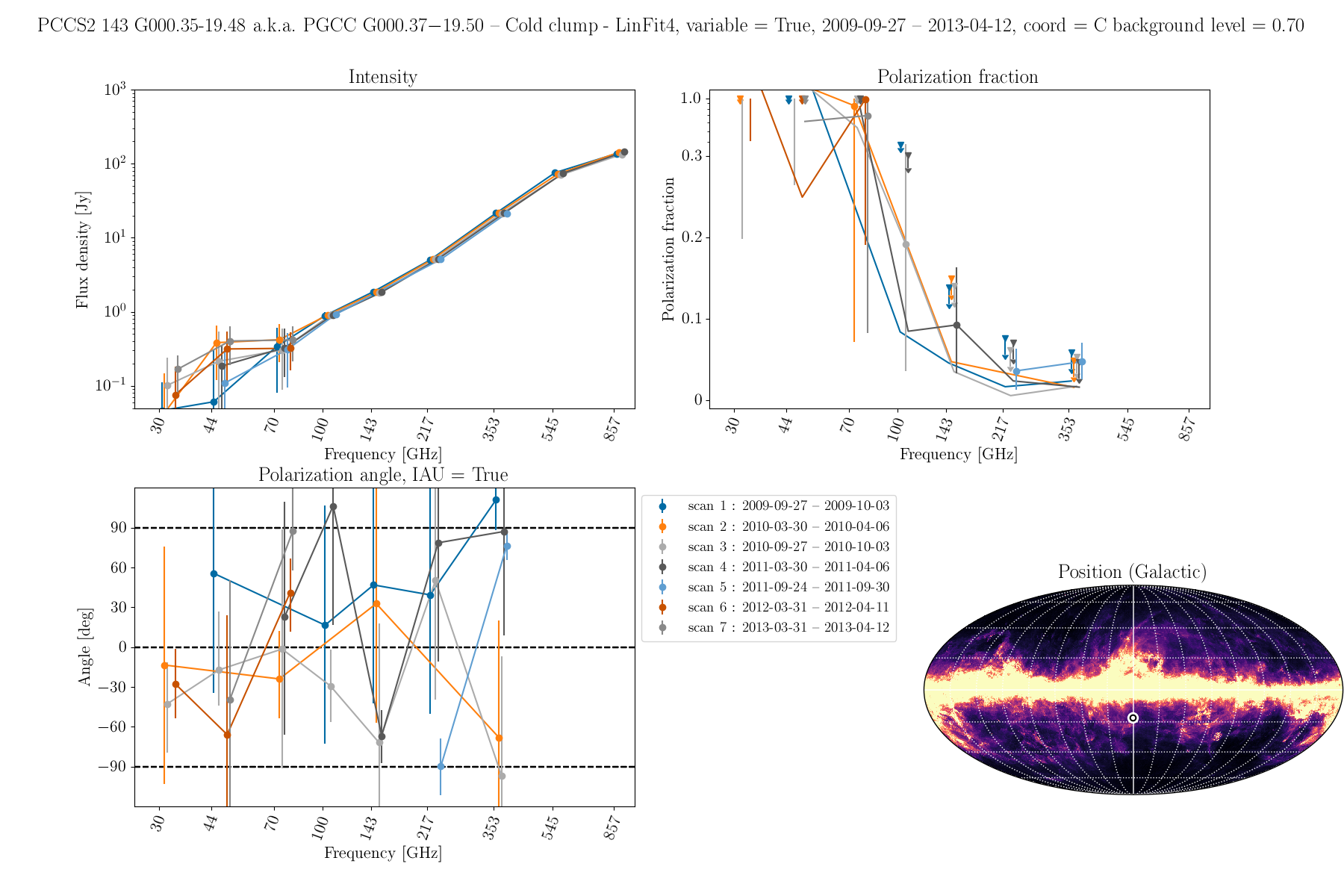}
    \caption{SED of a Galactic source, PCCS2\,143\,G000.35$-$19.48 at RA = $287.56\deg$ and dec = $-37.15\deg$.  The lines and points of different colors mark different observing epochs, separated by 6 months. It is important to note the larger scatter of the noisier LFI measurements at 30, 44, and 70\,GHz.  With the possible exception of 44\,GHz, there is no evidence for variability, as expected for a large source of this type.}
    \label{fig:Galac-source}
\end{figure}

\begin{figure*}
    \centering
    \includegraphics[trim=35 415 125 60,clip,width=\textwidth]{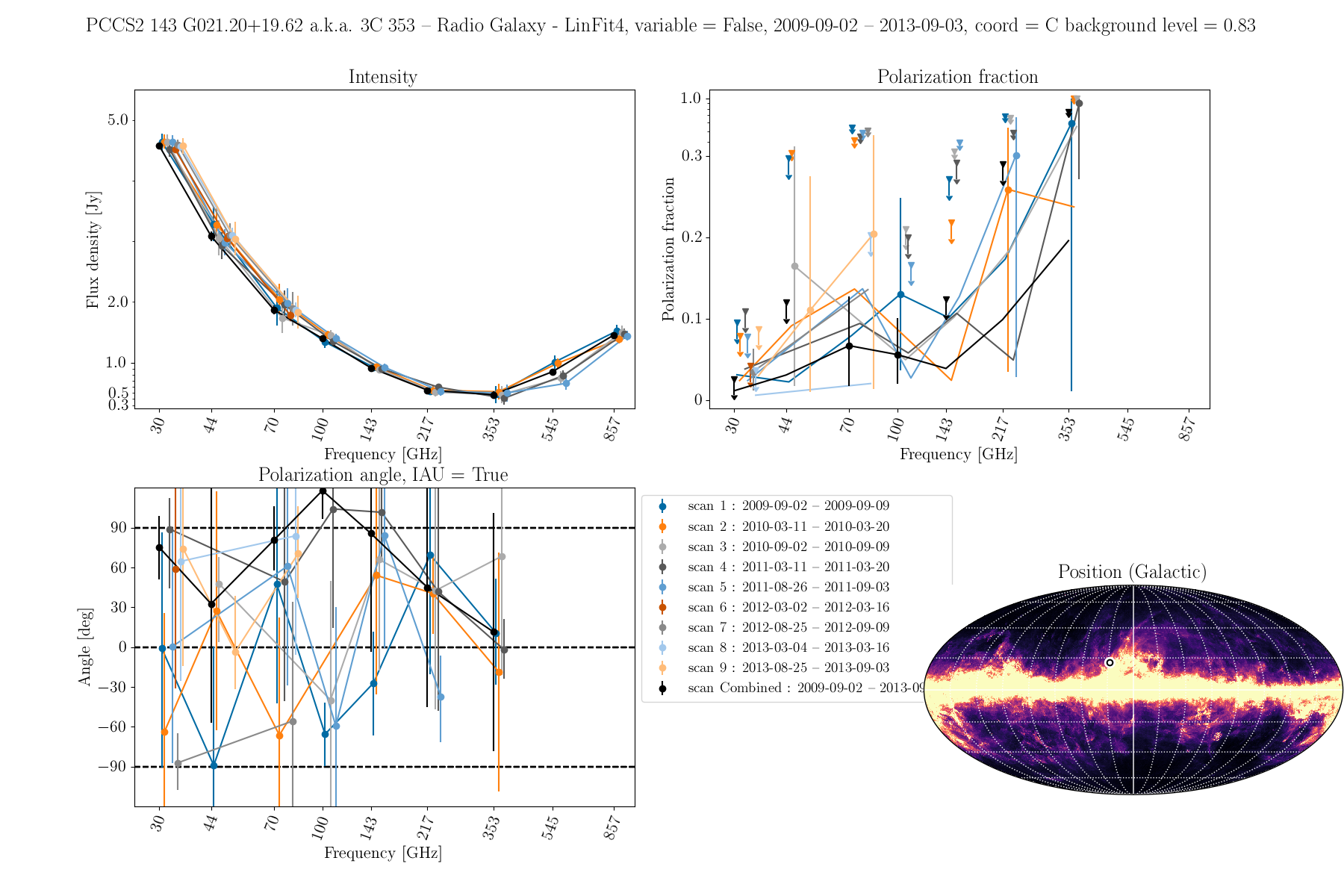}
    \includegraphics[trim=35 415 125 60,clip,width=\textwidth]{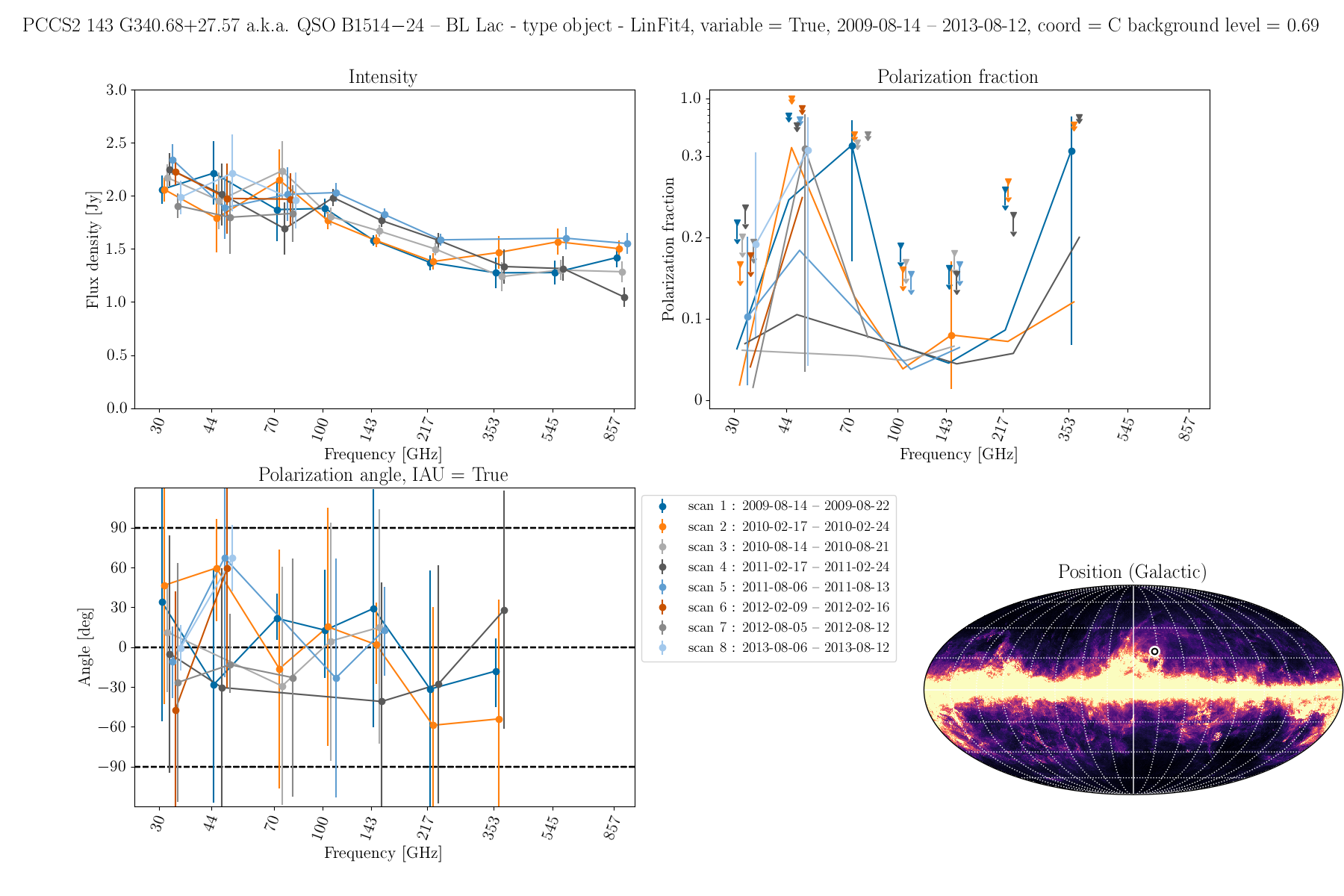}
    \caption{Same as in Fig.~\ref{fig:Galac-source}, but the lines and points of different colors mark different observing epochs, separated by 6 months.  For the polarization fraction (right column) we plot the maximum likelihood value as a solid line and indicate a 68\% confidence region or 95\% upper limit with error bars. Top: An AGN (3C\,353) displaying no significant changes in SED, and small, but apparently constant polarization.  Bottom: A BL\,Lac object, PKS\,1514$-$24, showing only mild variability, and no significant change in polarization.}
    \label{fig:nonvarsources}
\end{figure*}

The few Galactic (and LMC) sources are not beamed and have large physical dimensions, so they are unlikely to be variable on time scales of a few years or less.  They therefore offer an opportunity to check the fidelity of our pipeline.  Fig.~\ref{fig:Galac-source} provides an example. PCCS2\,143\,G000.35$-$19.48 is a Galactic dark cloud, and, as expected, shows no statistically significant variation with epoch at any \Planck\ frequency except for some scatter at 44\,GHz, which has the highest noise among the \Planck\ bands.  In addition, as shown in Fig.~\ref{fig:my_label}, the three 44\,GHz beams lie at the periphery of the \Planck\ field of view, and two of the beams have much higher ellipticity than the other one.  Consequently, the measured flux density at that frequency depends on how the source crosses the focal plane, and hence may show some instrument-related variation.  

Some extragalactic sources also demonstrate no significant variation during the \Planck\ mission.  We show two examples in Fig.~\ref{fig:nonvarsources}. G021.20+19.62 (3C\,353) is an radio galaxy with  a spectral upturn at 545\,GHz showing no variation in flux density at any frequency.  The polarization percentage, while small, also does not appear to vary.

Of more interest are those sources that clearly do vary during the \Planck\ mission. The blazar PCCS2\,143\,G086.11$-$38.19  (3C\,454.3), for instance, is a well-known and well-studied variable source that is known to have flared during the \Planck\ mission \citep{planck2011-6.2}.
We discuss 3C\,454.3 in some detail below (see Fig.~\ref{fig:3c454_vs_time}).

\begin{figure}
    \centering
    \includegraphics[trim=70 415 670 60,clip,width=\columnwidth]{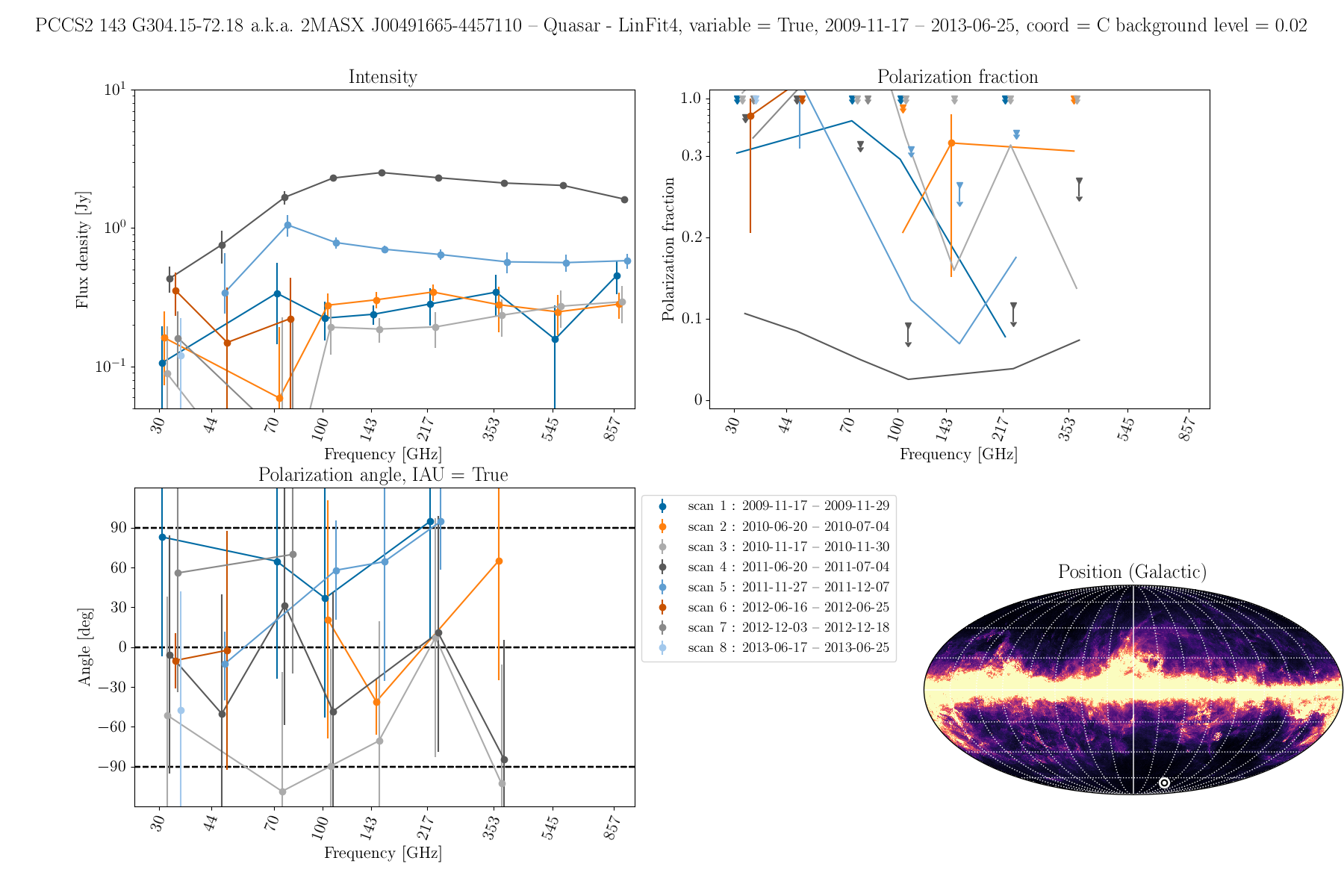}
    \caption{Strongly variable source, PCCS2\,143\,G304.15$-$72.18. It is apparent that the spectral shape, as well as luminosity, changes with epoch.}
    \label{fig:veryvarsources}
\end{figure}

In Fig.~\ref{fig:veryvarsources}, we show the SED of another strongly variable source, G304.15$-$72.18 at RA = $12.32\deg$, dec = $-44.95\deg$.  There is clear evidence of spectral change with epoch.  Strong variability is also evident in the following sources, among others: G106.95$-$50.62 (a Seyfert-1 galaxy), G315.80$-$36.52 (a QSO), and G352.45$-$08.41 (a blazar).

\subsection{Concurrent observations with other instruments}

\begin{figure}
    \centering
    \includegraphics[trim=20 160 50 100,clip,width=\columnwidth]{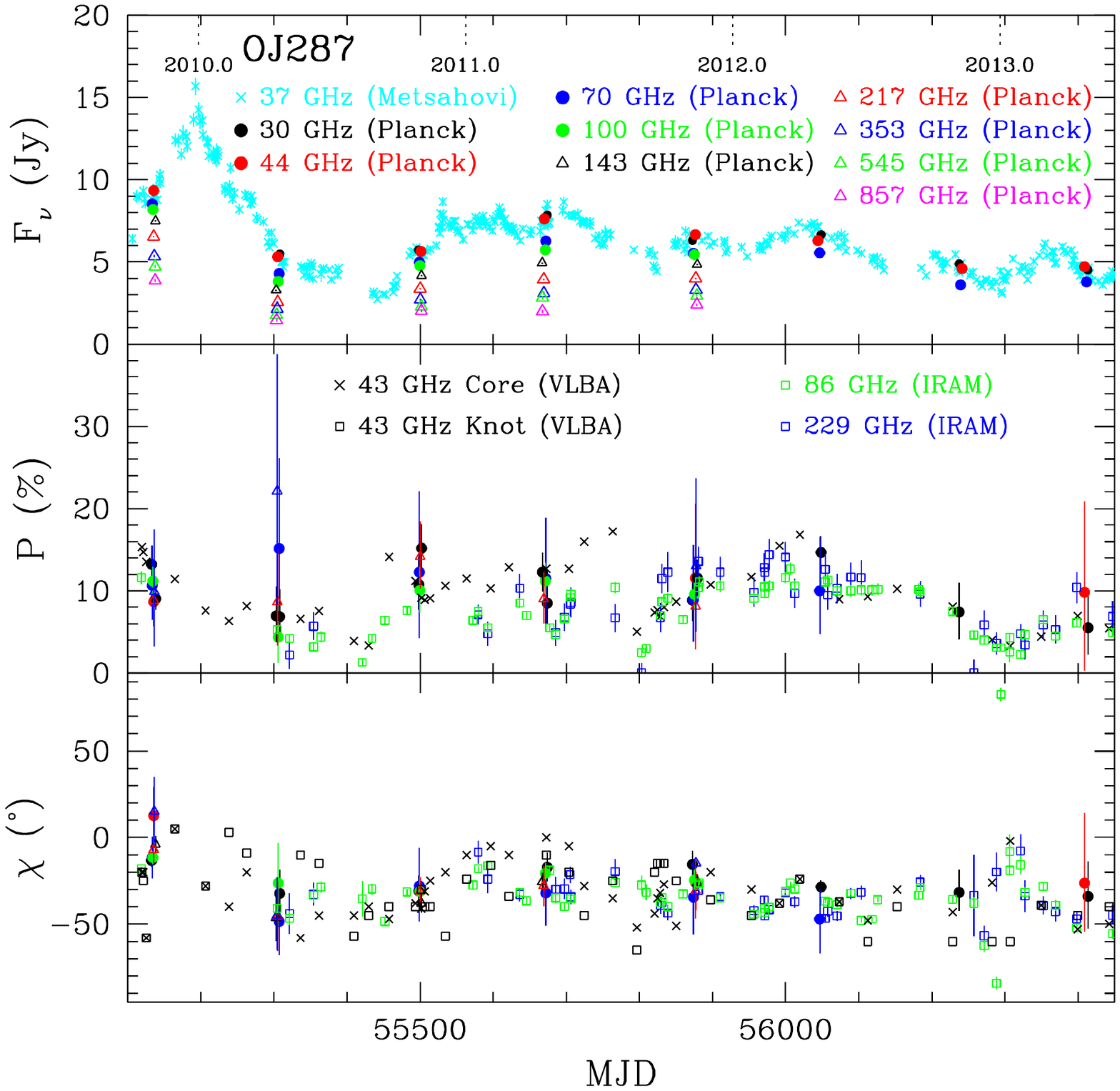}
    \caption{Flux density and polarization versus time of the BL~Lac object OJ\,287. Frequencies and sources of the data are indicated in the figure. The figure does not plot the following upper limits:
44\,GHz: MJD 55301--55309 ($\leq34\%$), 55497--55505 ($\leq30\%$), 55666--55674 ($\leq24\%$), 56038--56050 ($\leq27\%$), 56236--56247 ($\leq42\%$);
70\,GHz: MJD 56238--56241 ($\leq31\%$), 56410--56413 ($\leq31.2\%$);
353\,GHz: MJD 55304--55305 ($\leq77\%$),  55500--55501 ($\leq52\%$), 55669--55670 ($\leq37\%$).
No polarization was detected at 545 or 857\,GHz, since the HFI detectors are unpolarized at those frequencies.}
    \label{fig:oj287multi}
\end{figure}

\begin{figure}
    \centering
    \includegraphics[trim=20 160 50 100,clip,width=\columnwidth]{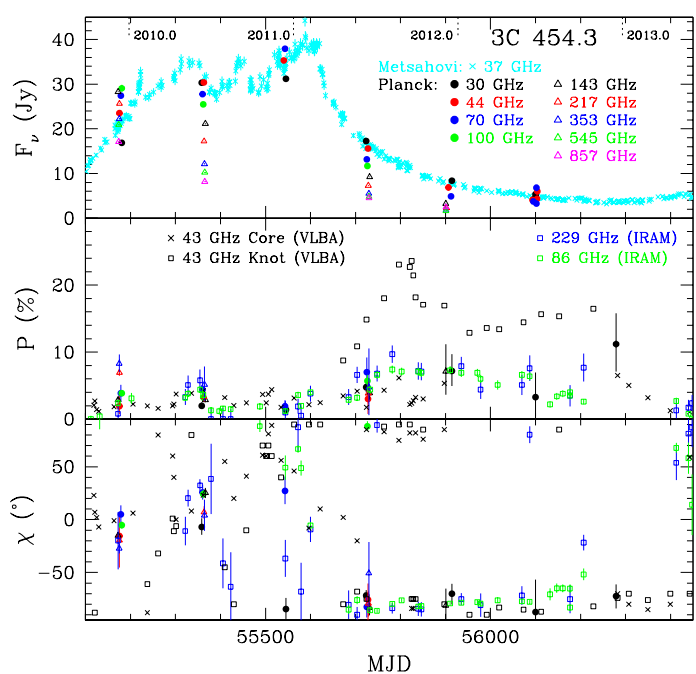}
    \caption{Flux density and polarization versus time of the quasar 3C\,454.3. Frequencies and sources of the data are indicated in the figure. The figure does not plot the following upper limits:
30\,GHz: MJD 55179--55182 ($\leq2.4\%$), 56464--56465 ($\leq11\%$);
44\,GHz: MJD 55357--55369 ($\leq3.6\%$), 55535--55546 ($\leq4.6\%$), 55899--55914 ($\leq36\%$), 56089--56097 ($\leq42\%$), 56277--56285 ($\leq41\%$), 56464--56472 ($\leq29\%$);
70\,GHz: MJD 55912--55913 ($\leq47\%$), 56101--56103 ($\leq64\%$), 56277--56279 ($\leq36\%$), 56465--56467 ($\leq18\%$);
217\,GHz: MJD 55902--55902 ($\leq27\%$);
353\,GHz: MJD 55901--55901 ($\leq100\%$).
No polarization was detected at 545 and 857\,GHz.}
    \label{fig:3c454multi}
\end{figure}

\begin{figure}
    \centering
    \includegraphics[width=\columnwidth]{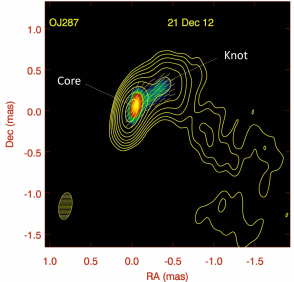}
    \caption{VLBA image of OJ\,287 at 43\,GHz at epoch 2012 December 21 (MJD 56282). The contours represent total intensity, starting at 0.25\% of the maximum of 0.79 Jy/beam and increasing by factors of 2. The color scale represents linearly polarized intensity, with maximum value (yellow) of 0.032 Jy/beam. The yellow line segments indicate the EVPA. The restoring beam FWHM, representing the angular resolution, is displayed in the lower left corner.}
    \label{fig:oj287vlba}
\end{figure}

\begin{figure}
    \centering
    \includegraphics[width=\columnwidth]{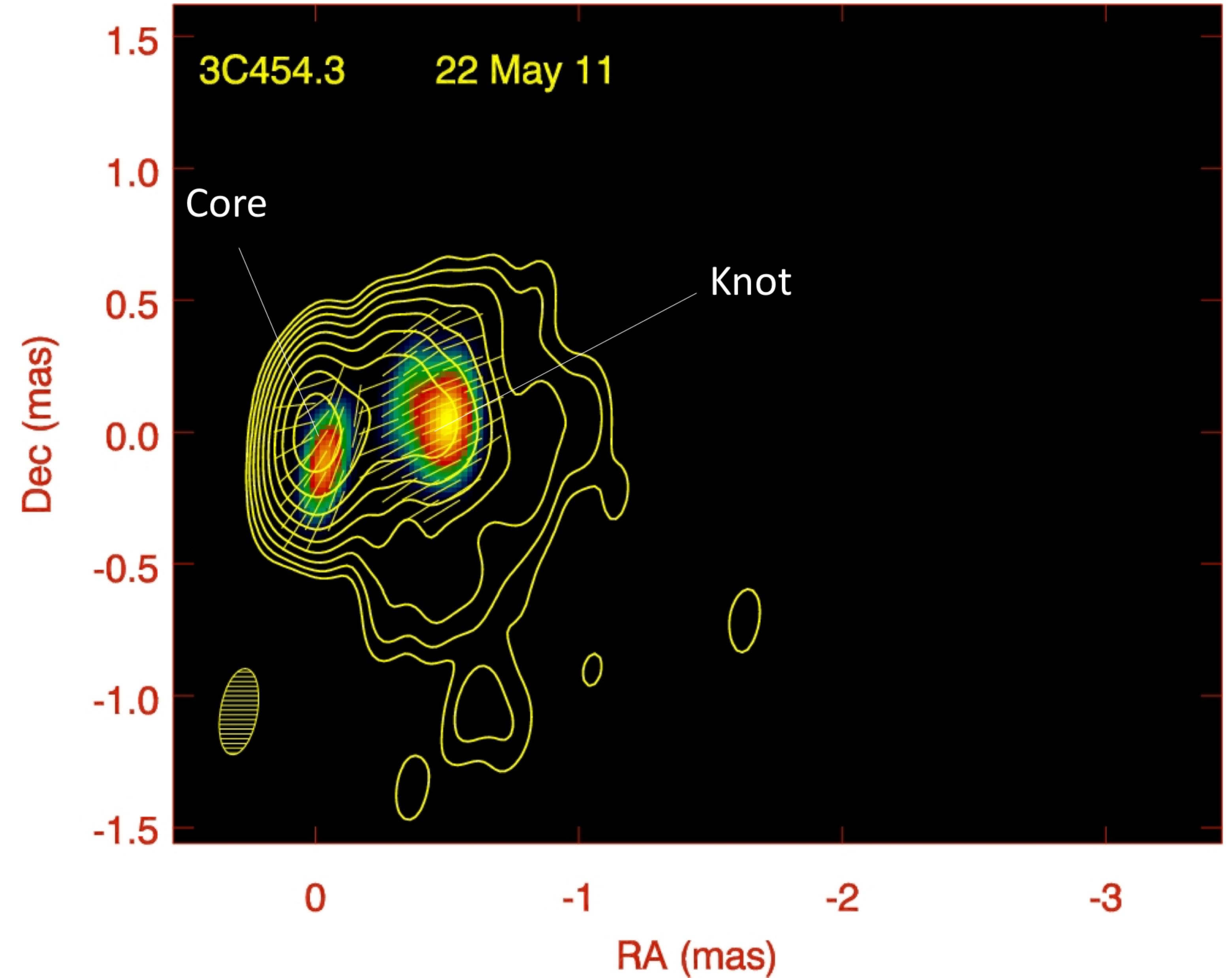}
    \caption{VLBA image of 3C\,454.3 at 43\,GHz at epoch 2011 May 22 (MJD 55704). The contours represent total intensity, starting at 0.25\% of the maximum of 6.00 Jy/beam and increasing by factors of 2. The color scale represents linearly polarized intensity, with maximum value (yellow) of 0.27 Jy/beam. The yellow line segments indicate the EVPA. The restoring beam FWHM, representing the angular resolution, is displayed in the lower left corner. }
    \label{fig:3c454vlba}
\end{figure}

In order to check our measurements and to illustrate possible uses of the data presented
here, we compare the \Planck\ polarization data of selected
objects with published data from other astronomical instruments.
Fig.~\ref{fig:oj287multi}
presents the \Planck\ data of the BL~Lac object OJ\,287 alongside the 37\,GHz light curve
from the Mets\"ahovi Radio Observatory (as summarized in 
\citealt{planck2016-XLV}),
linear polarization at 86 and 230\,GHz measured with the
30-meter antenna of the Institut de Radioastronomie Millim\'etrique
\citep[IRAM,][]{Agudo2018}, and Very Long Baseline Array (VLBA)
images that include linear polarization \citep{Jorstad2017}. Fig.~\ref{fig:3c454multi} provides a similar plot for the quasar 3C\,454.3.
For comparison, we present in Figures \ref{fig:oj287vlba} and \ref{fig:3c454vlba} VLBA images at 43\,GHz of OJ\,287 and 3C\ 454.3 obtained within one month of an epoch for which we have \Planck\ polarization measurements. The images are from the VLBA--BU Blazar Monitoring Program.\footnote{\label{fn_BU}\tt http://www.bu.edu/blazars/BEAM-ME.html}
We find that the flux densities at 30 and 44\,GHz agree with the Mets\"ahovi 37\,GHz values at nearby epochs. In addition, there is general correspondence between the \Planck\ and IRAM polarization. These similarities serve to validate our measurements.

\begin{figure}
    \centering
    \includegraphics[trim=20 150 40 100,clip,width=\columnwidth]{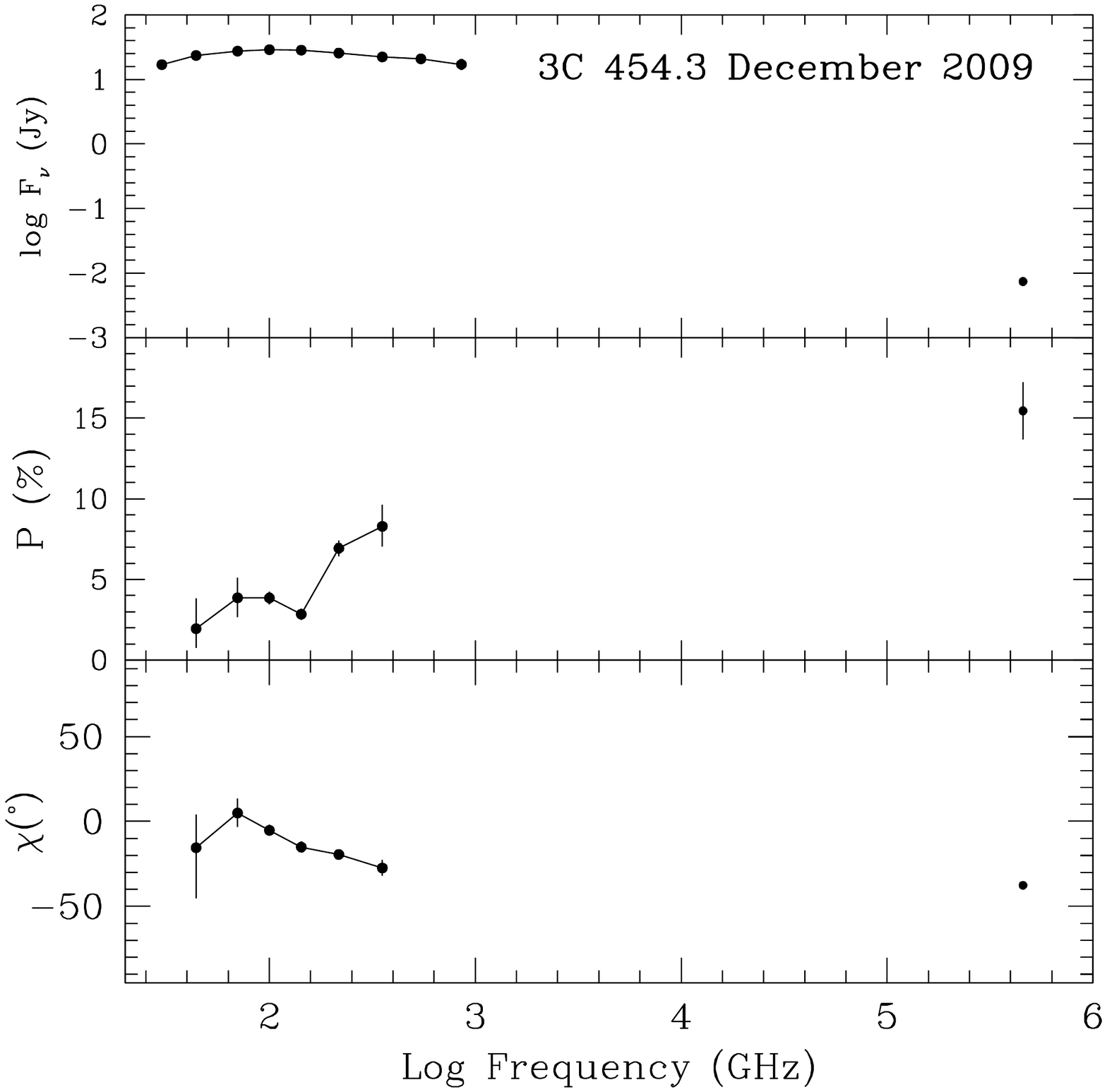}
    \caption{Frequency dependence of the flux density, degree of linear polarization, and EVPA of 3C\,454.3 in 2009 December. For frequencies $\leq353$\,GHz, the data are from this work. The high-frequency point is at optical R band from \citet{Jorstad2013}. }
    \label{fig:3c454epoch1}
\end{figure}
Figure \ref{fig:oj287multi} demonstrates that the \Planck\ polarization of OJ\,287 is similar to that of the core seen in Fig.~\ref{fig:3c454vlba}. We note that the optical electric-vector position angle (EVPA) of the knot deviates at some epochs from the source-integrated polarization at 30--353\,GHz.  \citet{Sasada2018} have shown that the optical polarization is related to that of either the core or the knot, depending on the epoch. Our data are consistent with this conclusion.  

Figure \ref{fig:3c454epoch1} presents the frequency dependence of the flux density, degree of polarization, and EVPA of 3C\,454.3 during the first \Planck\ survey in late 2009. At that epoch, an extremely high-amplitude two-year outburst was in its early, rising stage. The core completely dominated the polarization in a VLBA image obtained 17 days earlier (from the VLBA-BU-BLAZAR website). 
Figure \ref{fig:3c454epoch1} reveals that the degree of 
polarization was higher by a factor of $\sim2$ from 44--100 
GHz below the spectral turnover than at 217--353\,GHz above the
turnover. There is a frequency gradient in EVPA from
$\sim5^\circ$ to $\sim -30^\circ$ from 70 to 353\,GHz.
At the same epoch, the optical R-band EVPA was $-37^\circ\pm2.1^\circ$,
having shifted from $+25^\circ$ nine days earlier
\citep{Jorstad2013}. The degree of polarization at R band was $15.4\pm1.8\%$. This frequency dependence 
strongly implies that the outburst involved 
a component that dominated the emission from $\sim200$ 
GHz to optical frequencies. We are therefore able to conclude that the optical flare occurred within the same region as the $\sim200$\,GHz flare, which is distinct 
from the regions (probably farther downstream in the 
jet) that dominate the emission at lower frequencies.

\subsection{\Planck's contribution to the overall spectral energy distribution}

\begin{figure}
    \centering
     \includegraphics[width=\columnwidth,height=0.9\columnwidth]{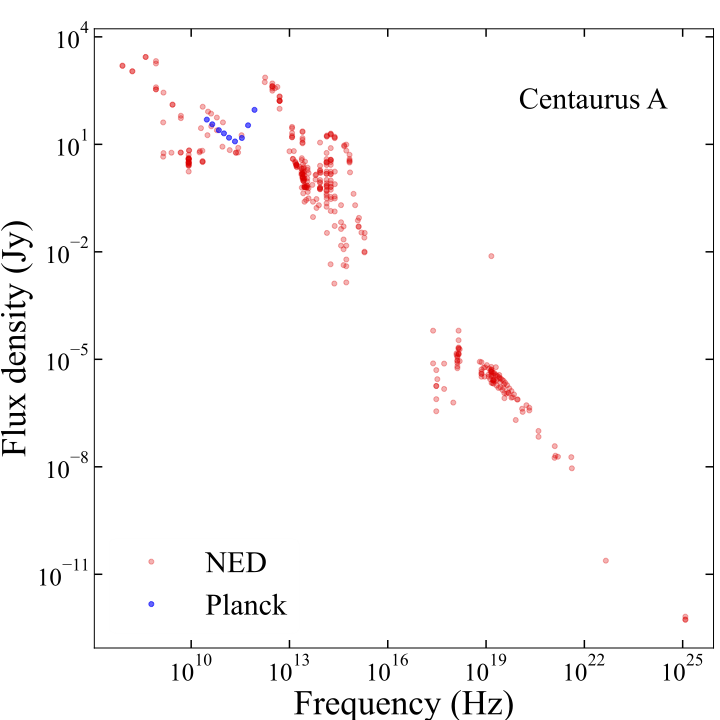}
     \includegraphics[width=\columnwidth,height=0.9\columnwidth]{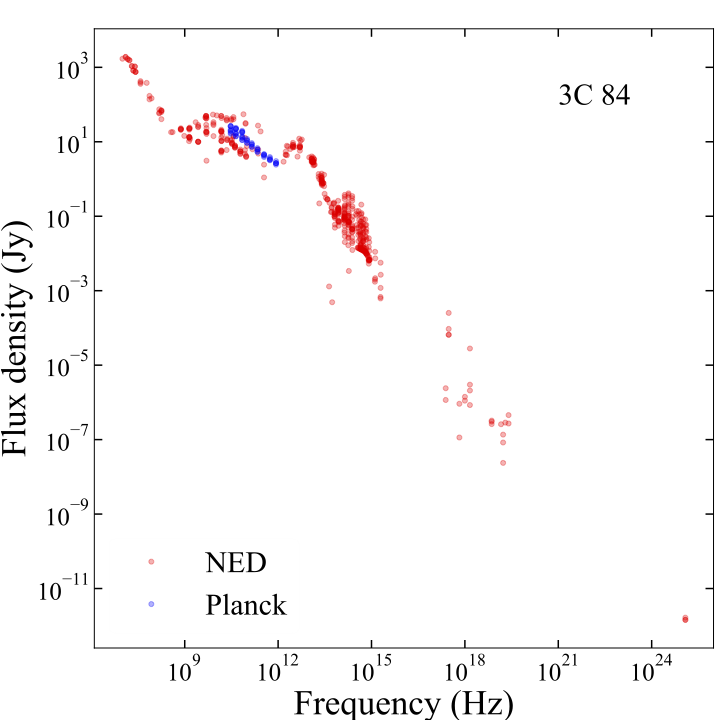}
    \includegraphics[width=\columnwidth,height=0.9\columnwidth]{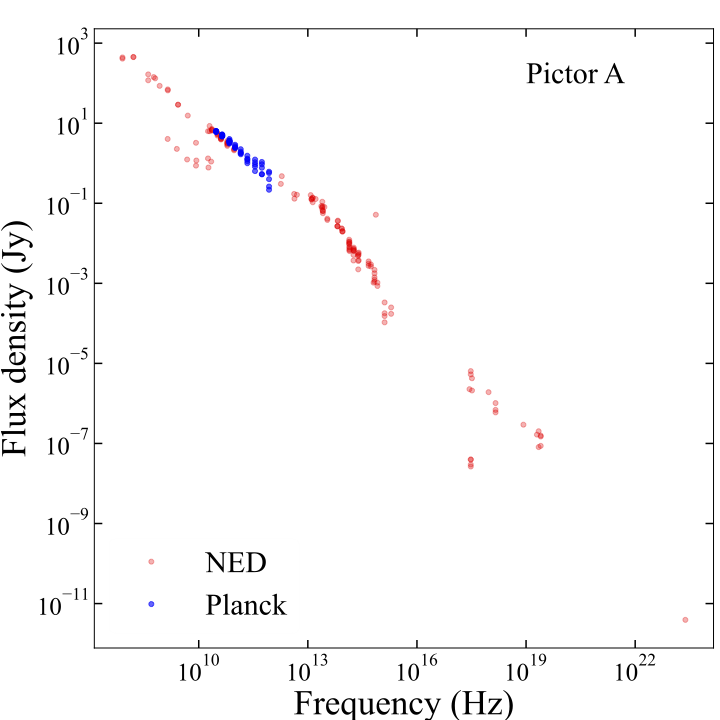}
    \caption{Spectral energy distribution (SED) for a few sources: red points are photometry from NED and the blue points are our \Planck\ observations. The \Planck\ data for Cen\,A add some new spectral coverage and help to define the rise of the dust component.}
    \label{fig:Co1}
\end{figure}

\begin{figure}
   \centering
        \includegraphics[width=\columnwidth,height=0.9\columnwidth]{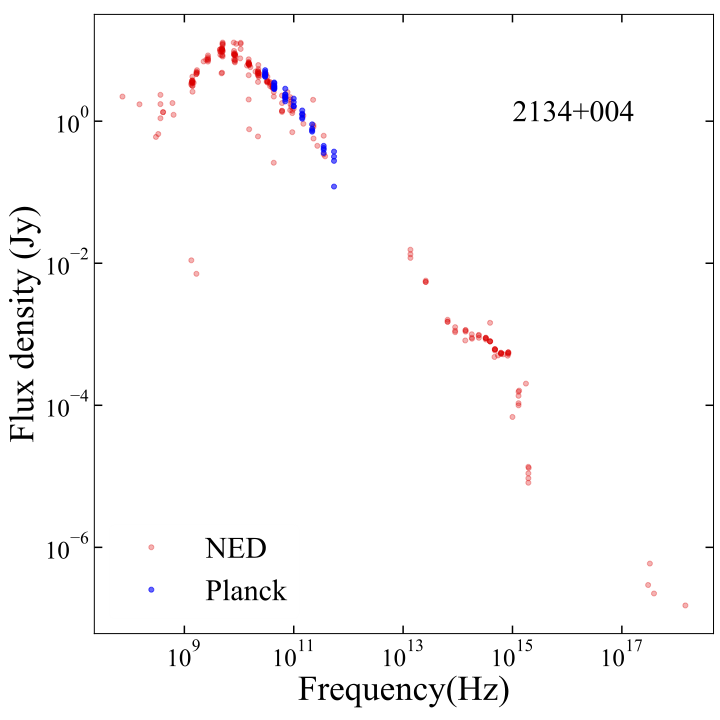}
       \includegraphics[width=\columnwidth,height=0.9\columnwidth]{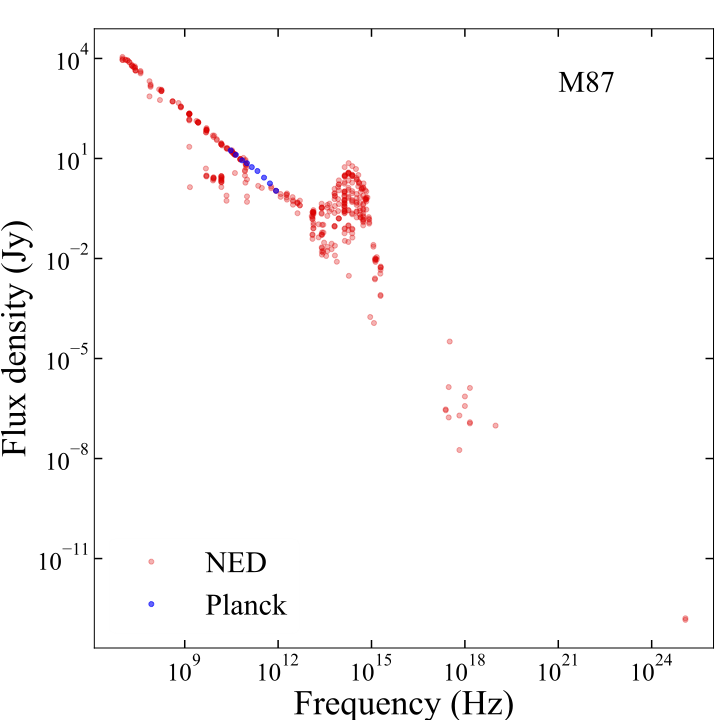}
    \caption{Spectral energy distribution (SED) for a few sources: red points are photometry from NED and the blue points are our \Planck\ observations. The \Planck\ observations of 2134+004 and M87 show good agreement with previous measurements.}
   \label{fig:Co2}
\end{figure}

In Figures \ref{fig:Co1} and \ref{fig:Co2} we present the \Planck\ data from PCCS2 \citep{planck2014-a35}.
For the remaining three sources, Centaurus\,A, 3C\,84, and Pictor\,A, the \Planck\ data help to fill in some missing spectral coverage.  In particular, in Centaurus\,A, the \Planck\ data help to define the rise into the spectral region dominated by the infrared dust component.

\section{Summary}
\label{sec:summary}

In this paper we have introduced the new \Planck\ Catalog of Polarized and Variable Compact Sources, PCCS-PV, comprising 153 sources, the majority of which are extragalactic, measured in both total power and polarization  by \Planck, with frequency coverage from 30 to 353 GHz, and time-scales ranging from days to years.
We classify 135 of the 153 sources as beamed, extragalactic radio sources (blazars), four as well-studied radio galaxies such as 3C\,84 and M87, and 14 as Galactic or Magellanic Cloud sources,  including \ion{H}{ii} regions, planetary nebulae, etc.
To allow an assessment of variability of polarized sources, the catalog was generated using an advanced extraction method, tod2flux, applied directly to the multifrequency
\Planck\ time-ordered data, rather than the mission sky maps. We used the calibrated timelines from the \Planck\ NPIPE data release, PR4,  as input to our processing.

To check our measurements and to illustrate possible uses of our catalog, we have compared the \Planck\ polarization data of selected objects with published data from other
astronomical instruments. 
Our preliminary findings show that for the selected sources OJ\,287 and 3C\,454.3, the flux densities at 30 and 44\,GHz agree with the Metsähovi 37\,GHz values at nearby epochs, while there is general correspondence between the \Planck\ and IRAM polarization. This general agreement confirms our measurements. 
Furthermore we have found that the \Planck\ polarization of OJ\,287 is similar to that of the core, supporting the conclusion of \citet{Sasada2018} that the optical polarization is related to that of the core at some epochs and the knot at others.
From joint analysis of our data and optical R-band data, we have shown a strong frequency dependency of the degree of polarization and EVPA of 3C\,454.3, with the EVPA at the highest \Planck\ frequencies becoming similar to the optical EVPA during the early stages of a major outburst. This has led us to conclude that the optical flare during the first \Planck\ survey in late 2009 occurred within the same region as the 200\,GHz flare, which is distinct from the regions that dominate the emission at lower frequencies.

We have also presented \Planck\ data along with observations compiled in NED for five non-blazar extragalactic sources to demonstrate \Planck’s contribution to the overall SED of the sources.
\Planck\ observations of M87 and the quasar 2134+004 are in good agreement with the observations in NED, confirming the flux density calibration of the \Planck\ data.
For the remaining three sources, Centaurus\,A, 3C\,84, and
Pictor\,A, the \Planck\ data help to fill in missing spectral coverage. In particular, in Centaurus\,A, the \Planck\ data help to define the rise as the infrared dust component starts to dominate the spectrum.

These preliminary results illustrate the usefulness of our \Planck\ catalog, PCCS-PV, for application with follow-up studies.
Furthermore, our versatile ToD-based flux-fitting software, tod2flux, is applicable to other experimental data sets.

PCCS-PV will be available at the ESA Planck Legacy Archive.\footnote{\url{https://www.cosmos.esa.int/web/planck/pla}} and at the NASA/IPAC Infrared Science Archive.\footnote{\url{https://irsa.ipac.caltech.edu/Missions/planck.html}}, while the flux-fitting algorithm, tod2flux, is released as open source software.\footnote{\url{https://github.com/hpc4cmb/tod2flux}}

\begin{acknowledgements}
The authors would like to acknowledge Charles Lawrence for insightful comments that helped improve this paper and Mainak Singha for his help with Figures 15 and 16.
This research was conducted under the auspices of a NASA Astrophysics and Data Analysis Program (ADAP) award NNH18ZDA001N-ADAP.

This work was carried out at the Jet Propulsion Laboratory, California Institute of Technology, under a contract with the National Aeronautics and Space Administration.\footnote{© 2021. All rights reserved. }

This study made use of Very Long Baseline Array (VLBA) data from the VLBA-BU-BLAZAR project, funded by NASA through the Fermi Guest Investigator Program. The VLBA is an instrument of the National Radio Astronomy Observatory. The National Radio Astronomy Observatory is a facility of the National Science Foundation operated by Associated Universities, Inc. The work at Boston University was supported in part by NASA through Fermi Guest Investigator Program grants 80NSSC17K0649 and 80NSSC20K1567.
C. O'Dea received support from the Natural Sciences and Engineering Research Council (NSERC) of Canada. 

This research has made use of the NASA/IPAC Extragalactic Database (NED),
which is operated by the Jet Propulsion Laboratory, California Institute of Technology,
under contract with the National Aeronautics and Space Administration.

This research used resources of the National Energy Research Scientific Computing Center (NERSC), a U.S. Department of Energy Office of Science User Facility located at Lawrence Berkeley National Laboratory, operated under Contract No. DE-AC02-05CH11231

This research has made use of the SIMBAD database \citep{2000A&AS..143....9W}, operated at CDS, Strasbourg, France.

\end{acknowledgements}

\section*{Software used in this work}

Astropy \citep{astropy:2013, astropy:2018},
HEALPix \citep{2005ApJ...622..759G},
Matplotlib \citep{Hunter:2007, thomas_a_caswell_2021_4743323},
Numpy \citep{harris2020array},
TOAST\footnote{\url{https://github.com/hpc4cmb/toast}}, and
Scipy \citep{2020SciPy-NMeth}.

\bibliographystyle{aa}
\bibliography{refs.bib}

\appendix

\section{Sample catalog file} \label{app:sample}

We show an example of a catalog file in Table \ref{tab:sample}. This one is for PCCS2\,143\,G009.33$-$19.61, better known as QSO B1921$-$293.  These results combine two adjacent frequency bands to boost the signal-to-noise ratio of each entry.

\begin{landscape}
\begin{table}
\begin{center}
\caption{Sample catalog file extract.}
\label{tab:sample}
\begingroup
    \fontsize{5pt}{10pt}\selectfont  
\begin{verbatim}
# Target = PCCS2 143 G009.33-19.61 a.k.a. QSO B1921-293 -- BL Lac - type object
# band(s)     , freq [GHz],      start,       stop, I flux [mJy], I error [mJy], Q flux [mJy], Q error [mJy], U flux [mJy], U error [mJy],    Pol.Frac,    PF low lim,  PF high lim, Pol.Ang [deg],   PA low lim,   PA high lim
       30+44,       32.182, 2009-09-29, 2009-10-06,      16000.0,          69.8,        -33.5,          94.8,       -136.1,          99.6,       0.0066,       0.0000,       0.0197,       -51.907,   -146.663,     42.720
       44+70,       58.818, 2009-10-04, 2009-10-05,      13443.3,         110.7,         51.9,         142.7,       -332.0,         175.5,       0.0225,       0.0009,       0.0485,       -40.555,    -70.395,     -0.150
      70+100,       96.610, 2009-10-03, 2009-10-04,      11271.7,          49.7,       -229.0,          74.7,         69.0,          73.1,       0.0202,       0.0086,       0.0339,        81.613,     63.813,     99.570
     100+143,      131.311, 2009-09-30, 2009-10-01,       9582.8,          27.0,       -117.7,          52.0,         83.7,          49.3,       0.0141,       0.0048,       0.0253,        72.301,     50.353,     92.881
     143+217,      167.344, 2009-10-02, 2009-10-03,       8157.3,          25.4,         55.0,          54.6,         60.4,          55.0,       0.0074,       0.0000,       0.0220,        23.858,    -84.555,    132.337
     217+353,      246.339, 2009-10-02, 2009-10-02,       6152.8,          39.5,         75.5,          69.5,       -102.6,          85.6,       0.0165,       0.0000,       0.0443,       -26.813,   -122.771,     69.173
     353+545,      430.560, 2009-10-01, 2009-10-02,       4169.6,          66.2,        315.2,         196.0,       -278.2,         182.7,       0.0907,       0.0102,       0.1924,       -20.717,    -53.489,      8.537
     545+857,      700.023, 2009-10-02, 2009-10-02,       2887.4,          70.8,          0.0,           0.0,          0.0,           0.0,       0.0000,       0.0000,       0.0000,         0.000,      0.000,      0.000
       30+44,       32.196, 2010-04-03, 2010-04-10,      15476.6,          68.4,       -539.3,          97.6,       -695.0,          95.4,       0.0565,       0.0448,       0.0686,       -63.906,    -69.335,    -58.331
       44+70,       57.373, 2010-04-04, 2010-04-05,      13108.5,         116.8,       -522.7,         153.6,       -745.5,         187.5,       0.0685,       0.0423,       0.0965,       -62.518,    -72.334,    -53.659
      70+100,       97.784, 2010-04-05, 2010-04-06,      10769.2,          46.0,       -488.5,          66.3,       -645.7,          71.4,       0.0749,       0.0634,       0.0872,       -63.553,    -68.112,    -59.239
     100+143,      129.550, 2010-04-08, 2010-04-09,       9214.9,          26.0,       -398.7,          45.5,       -507.1,          46.6,       0.0698,       0.0613,       0.0791,       -64.088,    -67.866,    -60.300
     143+217,      170.622, 2010-04-06, 2010-04-07,       7651.2,          24.1,       -301.8,          50.8,       -421.1,          49.5,       0.0674,       0.0560,       0.0794,       -62.816,    -68.104,    -57.499
     217+353,      244.762, 2010-04-07, 2010-04-07,       5886.6,          35.4,       -255.1,          76.6,       -396.1,          79.8,       0.0789,       0.0554,       0.1047,       -61.390,    -70.330,    -52.671
     353+545,      424.983, 2010-04-08, 2010-04-08,       4079.1,          63.5,        -94.1,         190.9,       -592.2,         197.1,       0.1393,       0.0560,       0.2371,       -49.517,    -68.426,    -31.301
     545+857,      719.622, 2010-04-08, 2010-04-08,       2799.4,          65.7,          0.0,           0.0,          0.0,           0.0,       0.0000,       0.0000,       0.0000,         0.000,      0.000,      0.000
       30+44,       32.800, 2010-09-29, 2010-10-06,      13487.8,          74.0,       -808.2,         100.7,       -908.7,         107.2,       0.0898,       0.0758,       0.1045,       -65.825,    -70.246,    -61.398
       44+70,       56.334, 2010-10-04, 2010-10-05,      10946.6,         117.3,       -550.9,         147.2,       -754.9,         192.6,       0.0839,       0.0569,       0.1142,       -63.060,    -73.898,    -54.073
      70+100,       97.415, 2010-10-03, 2010-10-04,       8374.8,          50.4,       -403.7,          76.9,       -470.8,          76.1,       0.0734,       0.0584,       0.0898,       -65.307,    -72.235,    -58.341
     100+143,      131.224, 2010-09-30, 2010-10-01,       6723.2,          27.1,       -375.2,          48.9,       -375.2,          47.2,       0.0786,       0.0663,       0.0917,       -67.499,    -72.366,    -62.452
     143+217,      171.378, 2010-10-02, 2010-10-03,       5538.4,          24.7,       -285.7,          50.1,       -328.1,          49.9,       0.0780,       0.0620,       0.0951,       -65.524,    -71.573,    -59.446
     217+353,      247.381, 2010-10-02, 2010-10-02,       4225.1,          35.2,       -145.5,          70.3,       -266.3,          79.1,       0.0697,       0.0381,       0.1058,       -59.323,    -73.573,    -46.672
     353+545,      434.882, 2010-10-01, 2010-10-02,       2879.9,          58.0,       -216.3,         209.6,         -7.7,         186.9,       0.0376,       0.0000,       0.1939,       -88.981,   -191.752,     13.857
     545+857,      676.404, 2010-10-01, 2010-10-02,       2002.6,          62.2,          0.0,           0.0,          0.0,           0.0,       0.0000,       0.0000,       0.0000,         0.000,      0.000,      0.000
       30+44,       32.419, 2011-04-03, 2011-04-10,      14244.5,          69.4,        326.6,          97.0,       -838.6,          97.8,       0.0628,       0.0510,       0.0755,       -34.361,    -40.128,    -28.505
       44+70,       58.093, 2011-04-04, 2011-04-05,      11569.0,         111.9,        200.3,         148.0,       -538.9,         170.4,       0.0478,       0.0235,       0.0766,       -34.807,    -48.958,    -17.658
      70+100,       97.611, 2011-04-05, 2011-04-06,       9406.2,          42.7,         66.5,          60.5,       -343.7,          69.8,       0.0367,       0.0231,       0.0512,       -39.524,    -49.208,    -30.093
     100+143,      128.005, 2011-04-08, 2011-04-09,       7908.3,          25.8,        106.8,          41.4,       -314.2,          46.4,       0.0416,       0.0310,       0.0530,       -35.615,    -42.292,    -28.686
     143+217,      174.305, 2011-04-06, 2011-04-07,       6387.1,          24.0,         67.5,          45.6,       -311.5,          48.3,       0.0494,       0.0362,       0.0638,       -38.890,    -46.535,    -30.882
     217+353,      242.701, 2011-04-07, 2011-04-07,       5053.1,          33.7,          8.8,          70.0,       -235.7,          71.2,       0.0446,       0.0199,       0.0732,       -43.936,    -61.493,    -27.401
     353+545,      425.582, 2011-04-08, 2011-04-08,       3366.3,          60.2,        170.5,         192.9,         60.8,         166.7,       0.0000,       0.0000,       0.1976,         9.819,   -130.564,    150.214
     545+857,      731.062, 2011-04-08, 2011-04-08,       2210.1,          60.8,          0.0,           0.0,          0.0,           0.0,       0.0000,       0.0000,       0.0000,         0.000,      0.000,      0.000
       30+44,       32.976, 2011-09-26, 2011-10-05,      14861.0,          61.4,        221.1,          77.2,      -1337.1,         101.0,       0.0911,       0.0786,       0.1037,       -40.305,    -43.258,    -37.292
       44+70,       55.448, 2011-10-02, 2011-10-04,      12614.1,          98.2,         -3.7,         128.8,       -966.9,         170.6,       0.0760,       0.0516,       0.1024,       -45.108,    -51.725,    -36.989
      70+100,       98.052, 2011-10-01, 2011-10-02,      10435.2,          39.5,       -165.8,          56.8,       -690.4,          61.8,       0.0678,       0.0572,       0.0790,       -51.753,    -55.985,    -47.497
     100+143,      131.682, 2011-09-27, 2011-09-28,       8835.1,          20.1,       -105.1,          36.7,       -588.2,          39.6,       0.0675,       0.0594,       0.0759,       -50.066,    -53.355,    -46.822
     143+217,      167.716, 2011-09-30, 2011-10-01,       7504.9,          18.6,        -88.0,          40.5,       -457.2,          43.0,       0.0618,       0.0515,       0.0728,       -50.446,    -54.913,    -45.827
     217+353,      240.365, 2011-09-29, 2011-09-30,       5802.6,          29.6,       -121.8,          61.4,       -253.7,          68.7,       0.0474,       0.0258,       0.0714,       -57.825,    -70.148,    -45.794
     353+545,      424.828, 2011-09-29, 2011-09-29,       4059.3,          57.1,       -166.5,         160.6,        -46.2,         174.9,       0.0027,       0.0000,       0.1124,       -82.245,   -222.468,     57.955
     545+857,      726.542, 2011-09-29, 2011-09-29,       2705.3,          57.7,          0.0,           0.0,          0.0,           0.0,       0.0000,       0.0000,       0.0000,         0.000,      0.000,      0.000
       30+44,       31.721, 2012-04-05, 2012-04-15,      12390.7,          51.6,        394.2,          71.0,      -1167.3,          73.3,       0.0993,       0.0888,       0.1100,       -35.671,    -38.737,    -32.585
       44+70,       58.491, 2012-04-06, 2012-04-08,       9759.9,          96.0,         31.9,         120.7,       -726.8,         155.4,       0.0735,       0.0452,       0.1037,       -43.743,    -53.242,    -34.629
       30+44,       30.997, 2012-10-03, 2012-10-05,      12778.7,          66.2,        909.7,          84.5,       -899.2,         104.7,       0.0998,       0.0861,       0.1143,       -22.335,    -26.058,    -18.493
       44+70,       63.664, 2012-10-01, 2012-10-03,       9027.2,         119.1,        732.3,         162.2,       -319.6,         195.3,       0.0855,       0.0587,       0.1182,       -11.788,    -25.542,      1.262
       30+44,       33.001, 2013-04-05, 2013-04-16,       7308.8,          57.0,          0.3,          78.1,       -305.9,          83.8,       0.0405,       0.0204,       0.0633,       -44.971,    -60.112,    -31.039
       44+70,       56.412, 2013-04-07, 2013-04-09,       5726.6,          88.7,       -104.8,         112.4,        152.7,         142.4,       0.0241,       0.0000,       0.0741,        62.229,    -44.755,    168.972
\end{verbatim}
\endgroup
\end{center}
\end{table}
\end{landscape}

\end{document}